\documentclass[preprint,floats,aps,12pt]{revtex4}
\usepackage{times} 
\usepackage{amssymb,amsmath}
\ifx\pdfoutput\undefined
\usepackage[usenames,dvips]{color}
\usepackage{graphicx}     
\else
\usepackage{epstopdf}
\usepackage[pdftex]{graphicx}
\usepackage[pdftex]{hyperref}
\hypersetup{a4paper,
    pdftitle={Manuscript on XXX} 
    pdfauthor={Uwe Thiele},  
pdfproducer={lateX},
pdfview=FitV,       
pdfstartview=FitB,
linkcolor=blue,     
citecolor=blue,     
pagecolor=blue,     
urlcolor=red,       
breaklinks=true,    
colorlinks=true,
citebordercolor=0 0 0,  
filebordercolor=0 0 0,
linkbordercolor=0 0 0,
menubordercolor=0 0 0,
pagebordercolor=0 0 0,
urlbordercolor=0 0 0,
pdfhighlight=/I,
pdfborder=0 0 0,   
backref=false,
pagebackref=false,
bookmarks=true,
bookmarksopen=true,
bookmarksnumbered=true
}
\fi
\ifx\pdfoutput\undefined
\DeclareGraphicsExtensions{.eps}
\else
\DeclareGraphicsExtensions{.jpg, .pdf, .tif}
\fi
\graphicspath{{.},{./Fig/}} 
\newcommand{\no}[1]{}
\newcommand{\mylab}[1]{\label{#1}}
\begin{document}
%
\title{Emergence of the bifurcation structure of a Langmuir-Blodgett transfer model}
\author{Michael H.~K\"opf}
\affiliation{D\'epartement de Physique, {\'E}cole Normale Sup\'erieure, CNRS, 24
rue Lhomond, 75005 Paris, France}
\email{michael.koepf@ens.fr}
\homepage{http://www.michaelkoepf.de}
\author{Uwe Thiele}
\affiliation{Institut f\"ur Theoretische Physik, Westf\"alische
 Wilhelms-Universit\"at M\"unster, Wilhelm Klemm Str.\ 9, D-48149 M\"unster, Germany}
\email{u.thiele@lboro.ac.uk}
\homepage{http://www.uwethiele.de}

\begin{abstract}
  We explore the bifurcation structure of a modified Cahn-Hilliard
  equation that describes a system that may undergo a first order
  phase transition and is kept permanently out of equilibrium by a
  lateral driving. This forms a simple model, e.g., for the deposition of
  stripe patterns of different phases of surfactant molecules through
  Langmuir-Blodgett transfer.
  Employing continuation techniques the bifurcation structure is
  numerically investigated employing the non-dimensional transfer velocity as
  the main control parameter.  It is found that the snaking structure of steady
  fronts states is intertwined with a large number of branches of time-periodic
  solutions that emerge from Hopf or period doubling bifurcations and end in
  global bifurcations (sniper and homoclinic). Overall the bifurcation diagram
  has a harp-like appearance.
  This is complemented by a two-parameter study (in non-dimensional
  transfer velocity and domain size) that elucidates through which
  local and global codimension 2 bifurcations the entire harp-like
  structure emerges.
\end{abstract}
%
%
\maketitle
%
%
%
\section{Introduction} \mylab{sec:intro}
%
Pattern formation that occurs at moving three-phase contact lines,
e.g., where a liquid recedes or advances on a solid substrate under a
gas atmosphere, poses significant challenges to both, the
experimentalist and the theoretician. A prominent example are
deposition patterns, like a ring-shaped coffee stain
\cite{Deeg2000pre}, that result from a dynamic process that involves
hydrodynamic flow of a solution or suspension, the dynamics of the
contact line and evaporation of the solvent
\cite{HaLi2012acie,Thie2014acis}. The problem is significantly
enriched by a surface activity of the solute, i.e., by the presence of
surfactants, as then Marangoni forces resulting from spatially
inhomogeneous surface tensions are induced by concentration gradients
\cite{CrMa2009rmp}. Although, the creation or avoidance of deposition
patterns is of high practical relevance \cite{HaLi2012acie} and
experiments show that a large number of systems produce a wide variety
of patterns \cite{Thie2014acis}, the understanding and control of this
entire class of pattern formation processes is not yet well
developed. Here we discuss a reduced model for the deposition of line
patterns of surfactant molecules on a plate that is withdrawn from a
water-filled trough covered by a floating surfactant monolayer and
numerically analyse its solution and bifurcation behaviour. This shall
allow us to better understand how the emergence of deposition patterns
in out-of-equilibrium situations is related to equilibrium phase
transitions that occur in the same system when considered without
driving.

A frequently used mathematical framework for problems where the
evolution of a free surface of a film or shallow droplet on a
substrate needs to be described is the long-wave (or lubrication)
approximation \cite{OrDB1997rmp}. Assuming that all relevant fields
have small gradients parallel to the solid substrate the governing
hydrodynamic transport equations and boundary conditions are
asymptotically reduced to a single (for single layers of simple
liquids \cite{Mitl1993jcis,ShKh1998prl,OrDB1997rmp,Thie2010jpcm}) or
coupled (for two or more layers of simple liquids, or various complex
liquids \cite{PBMT2005jcp,CrMa2009rmp,NaTh2010n,ThTL2013prl}) fourth
order, strongly nonlinear partial differential equations that describe
the evolution of film thickness profile(s) and pertinent concentration
field(s). Up to now only a few works investigate the deposition of
regular one-dimensional line patterns with such long-wave models
\cite{KGFC_Langmuir_10,FrAT2011prl,FrAT2012sm,DoGu2013el} - mainly
through direct time-simulations of the derived long-wave evolution
equations. Although this allows one to determine parameter regions
where line patterns may be deposited, the technique is less than ideal
when it comes to understanding the onset of patterning: The
time-simulation results support several hypotheses regarding the role
of local and global bifurcations in the patterning process (cf., in
particular, section 3.4 of Ref.~\cite{FrAT2012sm} and the concluding paragraph
of \cite{DoGu2013el}) but they provide at best an incomplete picture of
the bifurcation structure, in particular as they are not able to reveal unstable
solutions that might coexist with the stable solutions that correspond to the
deposition. A deeper analysis is asked for, if one wants to clearly identify
the involved bifurcation types and to understand the entire underlying
bifurcation structure.

\begin{figure}[tbh]
\includegraphics[width=0.9\hsize]{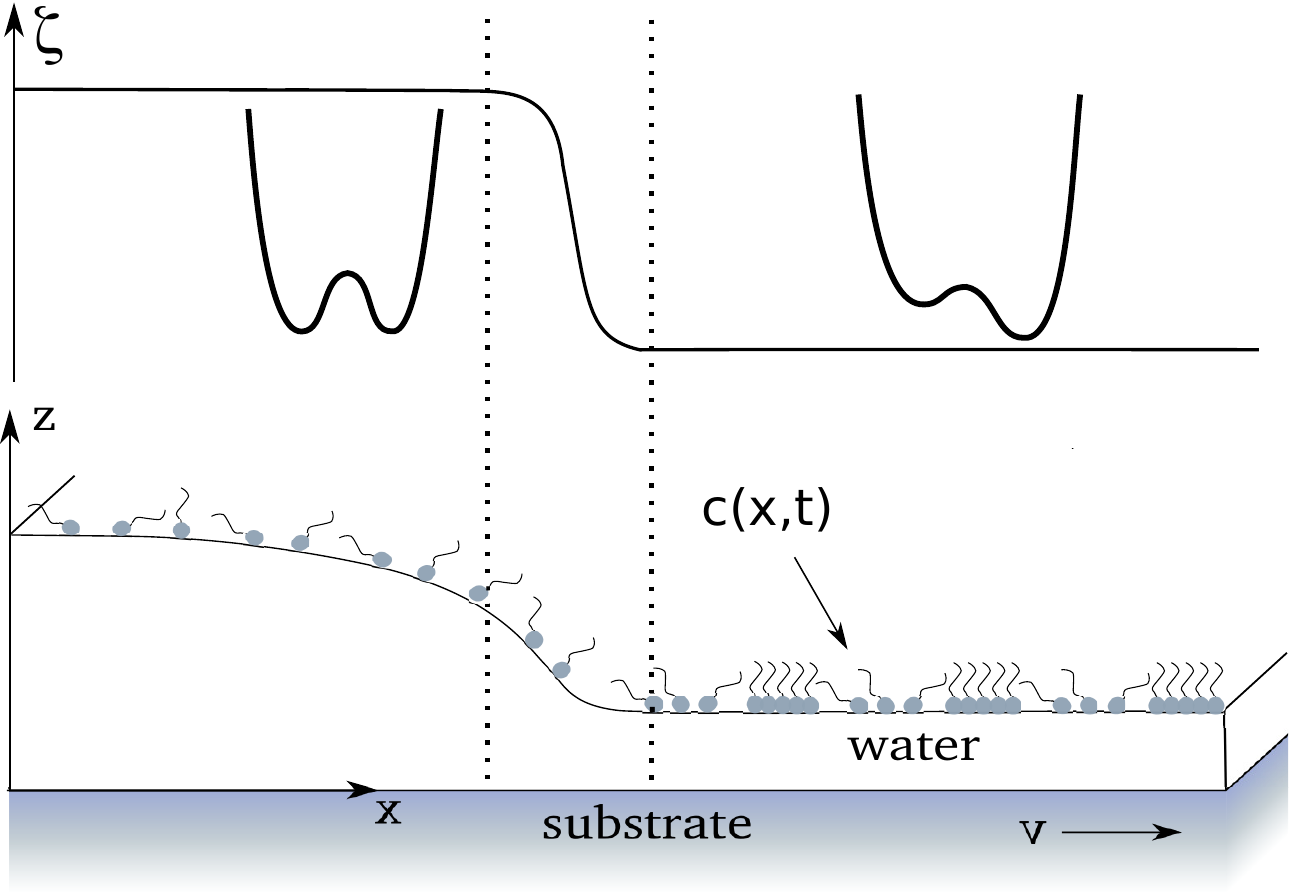}
\caption{Schematic drawing of a surfactant-covered meniscus on a solid
substrate during the transfer with velocity $v$. The surfactant density at the
surface is denoted by  $c(x, t)$. The function $\zeta(x)$ describes the
change of the tilt of the free energy from the symmetric situation far away from
the substrate (inset double well on the left) to the situation where the
condensed state is energetically favoured (asymmetric double well on the
right).)\mylab{fig:sketch}}
\end{figure}

Here, we focus on the Langmuir-Blodgett transfer of an insoluble
monolayer of surfactant from the surface of a water bath onto a plate
that is drawn out of the bath as sketched in Fig.~\ref{fig:sketch}. If
one uses a suitable amphiphilic molecule as surfactant at conditions
sufficiently close to a first order structural liquid-liquid phase
transition (liquid-expanded to liquid-condensed, see note \cite{note_SMC}) then
the intermolecular interaction with the solid plate may trigger a
phase transition in parts of the surfactant layer due to an effect
called substrate-mediated condensation
\cite{RS_ThinSolidFilms_92}. On a plate that moves out of a bath, the
partial condensation results in rather regular stripe patterns. They
are preferentially oriented either parallel or perpendicular to the
contact line. The wavelength of the resulting pattern depends on the
plate velocity and temperature and can range from several micrometers
down to a few hundred nanometers
\cite{GCF_Nature_00,KHRS_Langmuir_11,Li2012}.

For this system there exists a full long-wave description in terms of
coupled evolution equations for the film height profile and the
surfactant concentration \cite{KGF_EPL_09,KGFC_Langmuir_10} as well as
a reduced model in terms of a Cahn-Hilliard type evolution equation
for a density with amendments to take substrate-mediated condensation into account \cite{KGFT2012njp}.
Although the reduced model is a significantly simplified version of the
full model, it captures all essential properties including the instabilities
leading to stripe deposition parallel and perpendicular to the contact line and
even reproduces the results of the full model in the case of structured
substrates, where synchronisation effects can be observed
\cite{KGF_PRE_11,WiGu2013arxiv}. Most importantly, however, it allowed
for a first analysis of the bifurcations related to the onset of pattern
formation in the system. A preliminary bifurcation diagram has been obtained
(see Fig.~\ref{fig:numeration} below) by a combination of numerical
path continuation of steady states that correspond to homogeneous
deposition and direct numerical time simulation of time-periodic
solutions that correspond to the experimentally observed deposition of
regular line patterns.

It was found that a branch of stable time-periodic solutions exists for a
range of plate velocities $V$. It ends at high $V$ in a Hopf
bifurcation and at low $V$ in a homoclinic bifurcation. Furthermore,
it was possible to reveal a structure of the bifurcation diagram in
the range of low velocities which is best described by the term
\emph{front snaking} (see Fig.~\ref{fig:numeration} below). The branch
of steady solutions snakes back and forth, as more and more bumps are
added (one per wiggle) to a localized structure in the transition
region (i.e., front) between the bath surface and the drawn-out
film. As a result, several (stable or unstable) steady solutions can
exist at the same transfer velocity. This behaviour is reminiscent of
homoclinic snaking well known from the investigation of other model
equations of pattern formation, such as the standard Swift-Hohenberg
equation \cite{BK_PRE_06,BK_Chaos_07,Kno_Nonlinearity_08} and the
conserved Swift-Hohenberg equation \cite{TARG2013pre} (often called
phase-field crystal equation \cite{ELWG2012ap}).  However, as further
explained below, in our case, the solutions on the snaking curve
neither represent homoclinic nor heteroclinic orbits, but can be seen
as different front states.

The numerical approach of Ref.~\cite{KGFT2012njp} was limited w.r.t.\
the onset of patterning. It only permitted to find the loci of Hopf
bifurcations on the branch of steady solutions, but could not be
employed to determine the emerging branches of time-periodic solutions
from start to end. Therefore, important questions regarding the onset
of line pattern formation remain unanswered.
The present work overcomes these limitations employing numerical
continuation of stable and unstable branches of steady states (as in
Ref.~\cite{KGFT2012njp}) \textit{and} of time-periodic solutions. The
resulting completed bifurcation diagram in terms of our primary
control parameter -- the plate velocity $V$ -- is complemented by a
study of its dependence on a second dimensionless control parameter
that represents the ratio of the domain size to the typical patterning
length, i.e., is inverse to the 'undercooling' or quench strength. The
resulting two parameter study allows us to start to understand through
which codimension 2 local and global bifurcations the intertwined
structure of snaking steady states and time-periodic solutions
emerges.

The paper is structured as follows. Section~\ref{sec:model} briefly
reviews the geometric set-up and model equation of
Ref.~\cite{KGFT2012njp} that is employed here without change, followed
by a presentation of the employed numerical
method. 
Section~\ref{sec:res} presents our results while section~\ref{sec:conc}
concludes by discussing implications for systems that show similar
transitions and by pointing out further open questions and challenges.

%
\section{Model and numerical treatment} \mylab{sec:model}
\subsection{Evolution equation}
\mylab{sec:evolv}

The Langmuir-Blodgett transfer of surfactant monolayers close to a
first-order phase transition of the surfactant layer can lead to
regular stripe patterns. These stripes correspond to alternating
domains of different thermodynamic surfactant phases: liquid-expanded
phase (LE) and the liquid-condensed phase (LC) that can be seen as
respective analogues of the gas phase and the liquid phase in a
standard liquid-gas phase transition of a simple liquid.
The dynamics of the stripe formation can be modelled through a
long-wave approach that couples the sub-phase hydrodynamics to the
nonequilibrium thermodynamics of the monolayer via the surfactant
equation of state.  In the vicinity of a first-order phase transition,
this equation can be derived from a free energy functional of the type
proposed by Cahn and Hilliard~\cite{CaHi58} and determines the
dependence of the surface tension of the liquid-air interface on the
monolayer density, resulting in a Marangoni force and variations of
the Laplace pressure \cite{KGF_EPL_09,KGFC_Langmuir_10,ThAP2012pf}.
An analysis of the results of Ref.~\cite{KGFC_Langmuir_10} shows that
the full long-wave model may be simplified \cite{KGFT2012njp} and the
essence of the LE-LC domain formation under the influence of substrate-mediated condensation can
be captured by an amended Cahn-Hilliard-type model. The resulting
evolution equation consists of the generic Cahn-Hilliard equation
describing the dynamics of phase decomposition \cite{Cahn65} augmented
by the addition of a space-dependent external field and a dragging
term. The external field models the interaction between the monolayer
and the substrate that changes in a step-like manner in the contact
line region. The drag term describes that the contact line region
where the bath meets the moving plate recedes with nearly constant
speed in the reference frame of the plate.

We focus on the situation where the contact line is straight and the
plate is drawn in the direction orthogonal to the contact line. In
this setting the evolution of the scalar field $c(x,t)$ that represents
the concentration of surfactant on the surface of the liquid is captured by the
non-dimensionalised evolution equation \cite{KGFT2012njp}
\begin{equation}
\partial_t c = \partial_x \left[\partial_x\frac{\delta F}{\delta c}  - V c\right],
\label{eq:reduced}
\end{equation}
where $x$ is the spatial coordinate orthogonal to the
contact line (see Fig.~\ref{fig:sketch}), $t$ is time, and
$V$ is the speed at which the substrate is withdrawn.
The substrate dependent interaction field that causes substrate-mediated condensation enters through an explicite
space-dependent term
$\zeta(x)$ in the free energy density
\begin{equation}
f(c,x) = \frac{1}{2}(\partial_x c)^2 -\frac{c^2}{2} + \frac{c^4}{4} +
\mu \zeta(x) c. \label{eq:reducedf}
\end{equation}
It furthermore contains a gradient contribution that penalizes
interfaces and a double-well potential. One may say that $\zeta(x)$ is
responsible for a space-dependent tilt of the double-well potential.

The free energy density is integrated to obtain the total free energy $F=\int {\rm
d} x f(c,x)$.  We insert $F$ into Eq.~\eqref{eq:reduced} and obtain the 
evolution equation
\begin{equation}
\partial_t c = -\partial_x^2 \left[ \partial_x^2 c - c^3 + c - \mu
  \zeta(x) \right] - V\partial_x c.
\label{eq:reducedexpl}
\end{equation}
Note that due to the space-dependence of $f$, Eq.~\eqref{eq:reducedexpl} is not
invariant with respect to translations in $x$. Therefore one is not able to remove the
advection term $V \partial_x c$ by a simple Galilei transform. With other words, the
spatial dependence selects a particular frame of reference: the physical
laboratory system.

The function $\zeta$ smoothly switches between two thermodynamically different
regions: (i) left of the contact line ($x<x_s$, where $x_s$ is the contact line
position) the local free energy is a symmetric double well potential, i.e.\ the
liquid expanded (LE) and liquid condensed (LC) phases are energetically
equivalent, and (ii) right of the contact line ($x>x_s$) the local free energy
is tilted favouring the LC phase.  As the results are not very sensitive to the
particular form of $\zeta$ we use the same hyperbolic tangent centered at
$x=x_s$ as in Ref.~\cite{KGFT2012njp}:
\begin{equation}
\zeta(x) = - \frac{1}{2} \left[1 + \tanh \left(\frac{x-x_s}{l_s} \right)\right].
\end{equation}
Thus, the width of the transition region is determined by the constant
$l_s$. For consistency with Ref.~\cite{KGFT2012njp} we use the
boundary conditions (BC)
\begin{equation}
c(0) = c_0, \quad \partial_{xx}c(0) = \partial_x c(L) = \partial_{xx}c(L)=0. \label{eq:redbc}
\end{equation}
The density $c_0$ that is imposed at the left boundary of the system models the
presence of a virtually non-depletable surfactant bath. In fact, in the
experiments, the Langmuir trough is operated in a way that it keeps the
surfactant surface pressure, and therefore the density, constant by displacing barriers at the surface of the
trough.

Finally we mention a number of related models. Krekhov
\cite{Kr_PRE_09}, and Foard and Wagner \cite{FW_PRE_09,FW_PRE_12} use
related Cahn-Hilliard-type models. In particular, they employ space
dependent free energies that are switched from single-well potentials
to symmetric double-well potentials at a moving threshold, by
switching the sign of the quadratic term of the free energy. Such a
threshold models a quench front, dividing the spatial domain into a
two-phase and a one-phase region. This differs from the situation
considered here, as in our model the threshold divides two different
two-phase regions, namely, the region near the boundary on the
mensicus side where the free energy is a symmetric double well and the
region where the double well is tilted. The state in the region with
the symmetric double well is determined by the boundary condition.
The quench front investigated in
Refs.~\cite{Kr_PRE_09,FW_PRE_09,FW_PRE_12} propagates into a one-phase
region where the order parameter is spatially uniform. Thus, there is
a simple relation between the front velocity and the flow of material
to the quench front: If the value of the order parameter in this
region is called $c^\star$, then the flow of material to the front is
$c^\star V$ and remains constant with time.  In contrast to this
situation, the inflow of surfactant in our model is not constant but
dynamically adjusts to keep the surfactant density at the left
boundary constant. This reflects the physical reality of
Langmuir-Blodgett transfer, where the surfactant influx is determined
by the hydrodynamic flow field and the dynamics of the surfactant
molecules at the water surface, which in turn are influenced by the
Langmuir trough that keeps the surfactant density constant in the
bath.
%
\subsection{Non-dimensionalisation}
\mylab{sec:nondim}

Although for consistency, we use throughout this article the
non-dimensional equation \eqref{eq:reducedexpl} and non-dimensional
parameter as in Ref.~\cite{KGFT2012njp}, to understand the physical
role of the parameters it is instructive to review the underlying
scaling. We indicate dimensional quantities by a tilde and write the
dimensional equivalent of Eq.~(\ref{eq:reducedexpl}) as
\begin{equation}
\partial_{\tilde t} \tilde c = -M\partial_{\tilde x\tilde x}\left[
\sigma\partial_{\tilde x\tilde x}\tilde c
+\kappa(a\tilde c-\tilde c^3) - \tilde\mu\tilde\zeta(\tilde x)\right]
\,-\,\tilde{V} \partial_{\tilde x}\tilde c.
\label{eq:dimev}
\end{equation}
where $M$ is a diffusional mobility, $\sigma$ is related to interface
tension, $\kappa$ is an energy density scale, $\tilde\mu$ the chemical
potential difference and $\tilde V$ is the plate velocity.  Note, that
the concentration $\tilde c$ has no dimension as it is given as part
per volume. The parameter $a$ may be called a non-dimensional depth of
quench into the two phase region.

If one introduces the non-dimensional variables $c=  \tilde c/\Gamma$,
$ t = \tilde t/\tau$ and $x = \tilde x/\ell$ with scales
\begin{equation}
\Gamma=a^{1/2}, \qquad \tau=\frac{\sigma}{M a^2\kappa^2} \qquad\mbox{and}\qquad
\ell=\sqrt{\frac{\sigma}{a\kappa}}
\mylab{eq:scales}
\end{equation}
for concentration, time and length, respectively, one obtains the 
dimensionless Eq.~\eqref{eq:reducedexpl}. In this way the model
parameters are reduced to four: 
the non-dimensional chemical potential difference 
\begin{equation}
\mu=\frac{\tilde\mu}{a^{3/2}\kappa},
\mylab{eq:nondimnum1}
\end{equation}
the non-dimensional plate velocity
\begin{equation}
V=\frac{\tau}{\ell}\,\tilde V =\frac{\sigma^{1/2}}{M a^{3/2}\kappa^{3/2}}\,\tilde V ,
\mylab{eq:nondimnum2}
\end{equation}
the non-dimensional domain size $L=\tilde L / \ell$ and the scaled
concentration at the meniscus-side boundary $c_0$.
As the present study aims at understanding the onset of the deposition
of line patterns observed in Ref.~\cite{KGFT2012njp}, the same scaling
is used although other scalings exist that allow for distinct
parametric studies within the same physical parameter space
(see note \cite{note_scaling}).

The main focus here is the dependence of the solution behaviour on the
non-dimensional plate velocity that represents the primary control
parameter. If ones main interest is in the influence of temperature
(quench depth), as investigated in \cite{KHRS_Langmuir_11}, the
scaling sketched in the note \cite{note_scaling} is more
adequate. Here, the temperature dependent parameter $a$ appears in all
scales and dimensionless numbers \eqref{eq:scales} -
\eqref{eq:nondimnum2}.

The resulting bifurcation diagram in terms of our primary control
parameter shall be complemented by a study of transitions triggered by
changing a second parameter that represents a distance to
the threshold of structure formation at equilibrium. In the given
scaling, we chose the dimensionless domain size $L$, i.e., the ratio
of the physical domain size $\tilde{L}$ to the typical patterning length $\ell$. 
If one assumes the 'undercooling' $a$ and energy scale $\kappa$ are fixed,
changing $L$ amounts to a change of the interface tension $\sigma$.
We note, however, that it is difficult in any experimental realisation to
change $\sigma$ without changing $\kappa$ and $a$ as
well. Although, one could chose $\mu$ or $c_0$ as second parameter, we
think that this does not allow for such a clear physical
interpretation as does the choice of $V$ and $L$ as control
parameters. The two-parameter study in $V$ and $L$ presented in
section~\ref{sec:res} elucidates how the intertwined structure of snaking
steady states and time-periodic solutions emerges.
%
\subsection{Numerical approach}
\mylab{sec:num}
%
For the numerical continuation \cite{DoKK1991ijbc,Kuzn10,DWCD2014ccp}
of steady and time-periodic solutions of Eq.~(\ref{eq:reducedexpl})
with BC~(\ref{eq:redbc}), we employ spatial discretization onto an
equidistant grid and approximate the PDE by a large number of ODEs,
one for each grid point.  Here, we choose second order finite
differences to approximate spatial derivatives, that is, exactly the
same discretization scheme that has been used in the direct numerical
simulations in Ref.~\cite{KGFT2012njp}. Due to the simplicity of this
scheme, the boundary conditions of the problem can be easily
implemented what would, for example, not be the case with spectral
Chebychev methods.

As a result we obtain a dynamical system consisting of $n$ ODEs. It describes
the time evolution of the concentration values $c_i$ at equidistant points
$x_i$. We use the package auto07p \cite{DoKK1991ijbc,AUTO07P} to (i) perform
pseudo-arclength continuation \cite{Ke_Appl_77} of fixed points of the
dynamical system that correspond to steady solution profiles $c_0(x)$ of the
PDE; to (ii) detect saddle-node and Hopf bifurcations of the fixed points, and
most importantly (iii) to continue stable \textit{and} unstable time-periodic
solutions of the dynamical system that correspond to time-periodic solutions of
the PDE that represent the deposition of regular line patterns. Furthermore we
track (iv) the loci of saddle-node and Hopf bifurcations in an appropriate
two-parameter plane.

Note that the unstable time-periodic solutions are not accessible by
direct numerical integration since the numerical treatment necessarily
involves fluctuations. However, the structure of the bifurcation
diagram can only be completely understood if the unstable solutions
are included. Continuation has previously been employed for kinetic
equations like Eq.~(\ref{eq:reducedexpl}) and recent examples include
studies of depinning droplets in 1d \cite{ThKn2006njp} and 2d
\cite{BeTh2010sjads,BKHT2011pre}, of drawn menisci of simple liquids
\cite{GTLT14arxiv} and of finite-time sigularities in film rupture
\cite{TsBT2013ijam}. A brief overview is given in section~4.2 of
Ref.~\cite{DWCD2014ccp}.

The solutions of the model equations are vectors in a high dimensional space.
In order to visualize and further analyse the bifurcation behaviour of the
system, we define as solution measure the norm
\begin{equation}
||\delta c|| = \left[\frac{1}{TL}\int_0^T\int_0^L c(x,t)^2\, dx\, dt\right]^{1/2},
\end{equation}
where $L$ is the domain size and $T$ the temporal period for the
time-periodic solutions. For steady states and when
  investigating system trajectories, the time average
$(1/T)\int dt \ldots$ is omitted.
%
\section{Results} \mylab{sec:res}
%
\subsection{Steady profiles} \mylab{sec:steady-100}

\begin{figure}
\centering
\includegraphics[width=0.8\hsize]{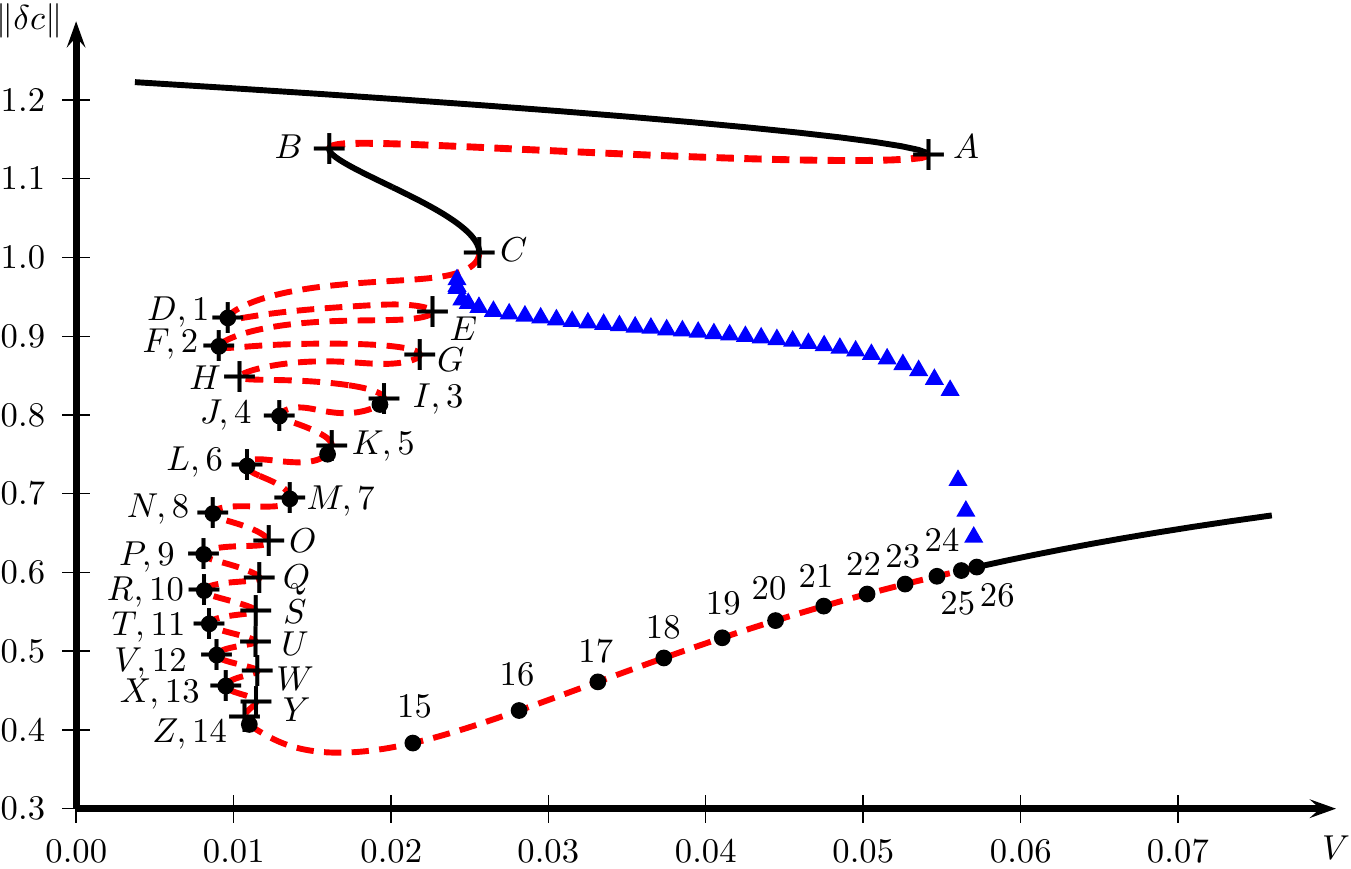}
\caption{Shown is the (time-averaged) $\lVert \delta c \rVert$ norm of steady and
  time-periodic solutions of Eq.~(\ref{eq:reducedexpl}) in dependence
  of the dimensionless plate velocity $V$ for $L=100$, $\mu=0.5$,
  $x_s=10$, $l_s=2$ and $c_0=-0.9$. The solid and dashed lines
  represent stable and unstable steady states (homogeneous deposition),
  respectively, as obtained by numerical path continuation. The
  blue triangles correspond to time-periodic solutions (line deposition)
  obtained by direct numerical simulation. Letters A--Z label
  saddle-node bifurcations (folds), while numbers $1$--$26$ label Hopf
  bifurcations (HB). Selected steady profiles are given in
  Fig.~\ref{fig:profiles}.}
\mylab{fig:numeration}
\end{figure}

\begin{figure} 
\centering 
\includegraphics[width=0.45\hsize]{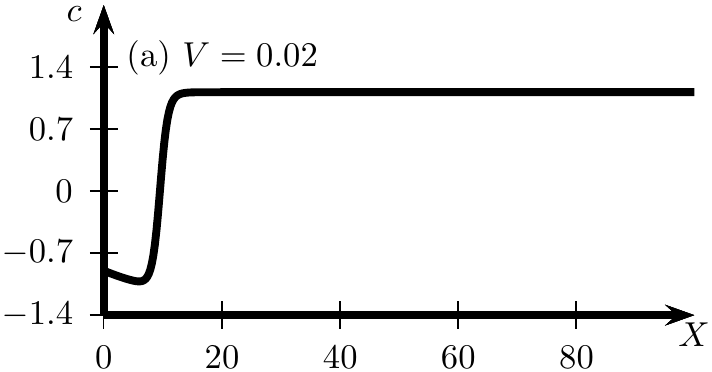}
\includegraphics[width=0.45\hsize]{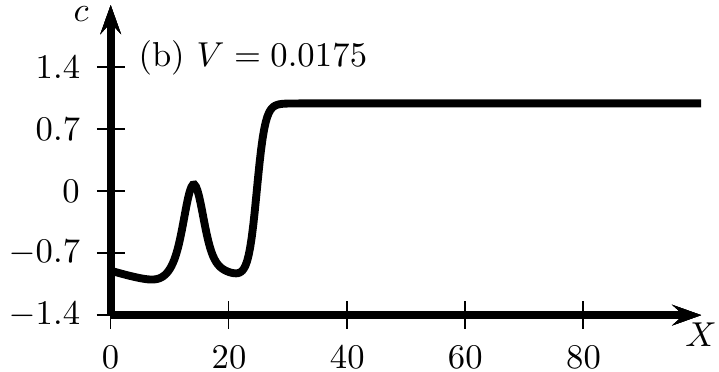} 
\includegraphics[width=0.45\hsize]{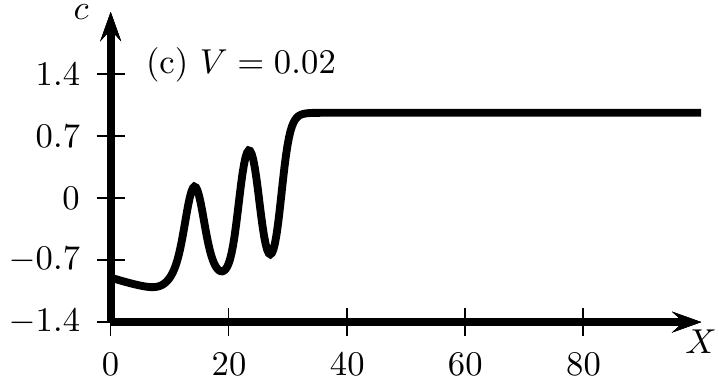}
\includegraphics[width=0.45\hsize]{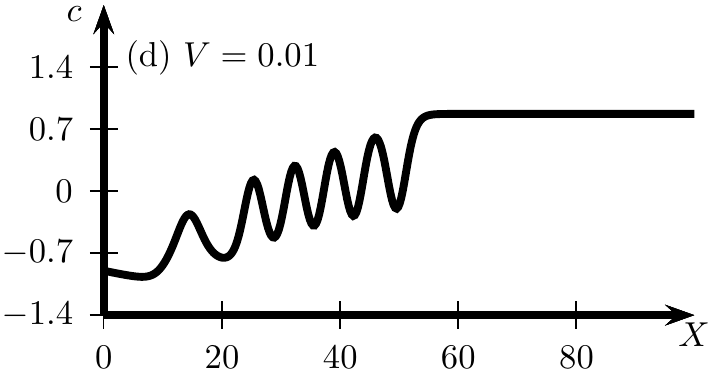} 
\includegraphics[width=0.45\hsize]{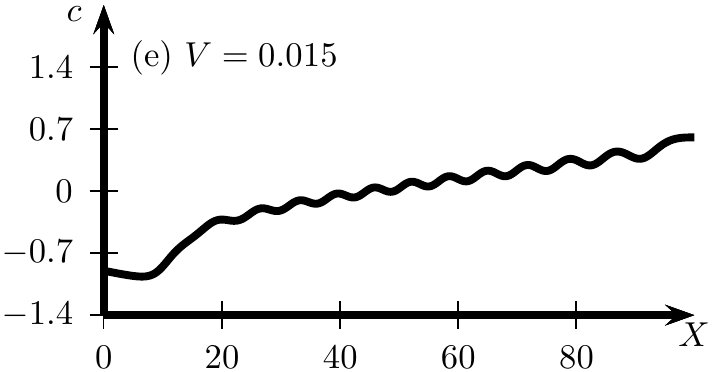}
\includegraphics[width=0.45\hsize]{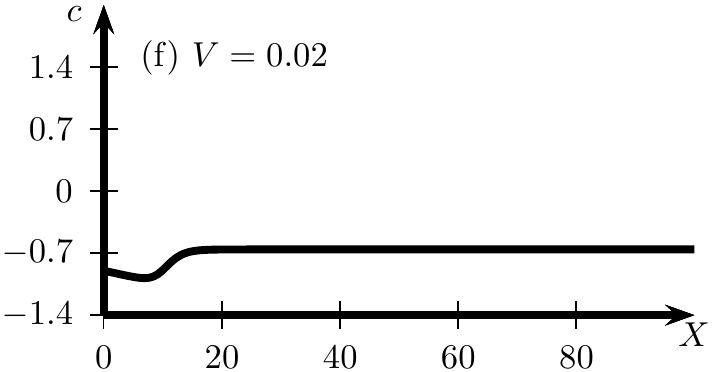} 
\caption{\mylab{fig:profiles} Examples of steady solution profiles
  from various points along the family of steady solutions in
  Fig.~\ref{fig:numeration}: (a) between the folds $B$ and $C$, (b)
  between the folds $D$ and $E$, (c) between the folds $G$ and $H$,
  (d) between the folds $L$ and $M$, (e) between the folds $Y$ and
  $Z$, (f) to the right of Hopf bifurcation 26.}
\end{figure}
In our previous investigation \cite{KGFT2012njp} a combination of
direct numerical simulation and numerical continuation of steady
solutions was used to construct a preliminary bifurcation diagram with
the plate velocity $V$ as primary control parameter, as summarised in
Fig.~\ref{fig:numeration}. The diagram includes a family consisting of
stable and unstable steady states (solid black and dashed red lines,
respectively), and a branch of stable time-periodic solutions (blue
triangles) that seems discontinuous close to its end at large $V$. The
parameters employed with Eq.~(\ref{eq:reducedexpl}) are $L=100$,
$\mu=0.5$, $x_s=10$, $l_s=2$ and $c_0=-0.9$. With the exception of $L$
these values are kept throughout the present study.  A linear
stability analysis of the steady solutions provides the loci of $26$
Hopf bifurcations (HB) that are numbered $1$--$26$ in
Fig.~\ref{fig:numeration}. The 26 saddle-node bifurcations (folds) of
the snaking curve are labeled by letters $A$--$Z$. Note that the various
saddle-node and Hopf bifurcations are not created by a few eigenvalues that
repeatedly cross the imaginary axis. Instead, nearly all of the Hopf
bifurcations that are situated on the steady state branch left of the point of
minimal norm (at $V_\mathrm{min}\approx0.015$) in Fig.~\ref{fig:numeration} are
related to different pairs of complex conjugate eigenvalues that aquire a
positive real part. All of them become stable again on the piece of steady
state branch right of $V_\mathrm{min}$, giving rise to more Hopf bifurcations.
The steady profile is linearly stable for $V\gtrsim0.06$ where the last HB
occurs. Therefore our further study focuses on the velocity range $0<V<0.1$
where all the mentioned bifurcations occur. Selected steady states profiles are
shown in Fig.~3.).

On the snaking structure one observes two basic scenarios for the relation of
saddle-node and Hopf bifurcations: In scenario (i) first a negative real
eigenvalue crosses zero at a saddle-node bifurcation rendering the
corresponding eigenmode unstable, then the same eigenvalue becomes negative
again at a second saddle-node bifurcation, briefly before forming a complex
conjugate pair together with another negative real eigenvalue. The complex
conjugate pair then crosses the imaginary axis at a Hopf bifurcation. Examples
for this scenario are all such sequences between from ($O, P, 9$) to ($Y, Z,
14$) in Fig.~\ref{fig:numeration}. As the second saddle-node and the Hopf
bifurcation are often very close to each other the system might be very close
to a codimension 2 Bogdanov-Takens bifurcation. In scenario (ii) first a
complex conjugate pair of eigenvalues with negative real part crosses the
imaginary axis at a Hopf bifurcation. The pair then splits into two positive
real eigenvalues. The smaller of the two becomes negative at a saddle-node
bifurcation before it joins another negative real eigenvalue to become another
complex conjugate pair of eigenvalues.  Briefly afterwards the pair crosses the
imaginary axis at a second Hopf bifurcation. This second Hopf bifurcation can
form the first Hopf bifurcation of the next such sequence.  Examples for this
scenario are all such sequences from ($3, J, 4$) to ($7, N, 8$) in
Fig.~\ref{fig:numeration}.

Next, in section~\ref{sec:res-100}, we complete the bifurcation diagram for
$L=100$, by a systematic continuation of all branches of time-periodic
solutions that emerge at the described Hopf bifurcations.  Then, in
section~\ref{sec:res-small}, we turn our attention to the snaking structure in
the range $0.005\lesssim V \lesssim 0.025$ of the bifurcation diagram. To
better understand the relation of the saddle-node and Hopf bifurcations of the
branch of steady solutions, we investigate how these structures and the
time-periodic branches emerge/vanish when changing the dimensionless domain
size $L$.
%
\subsection{Completed bifurcation diagram for $L=100$ \mylab{sec:res-100}}
%
\begin{figure}
\includegraphics[width=0.8\hsize]{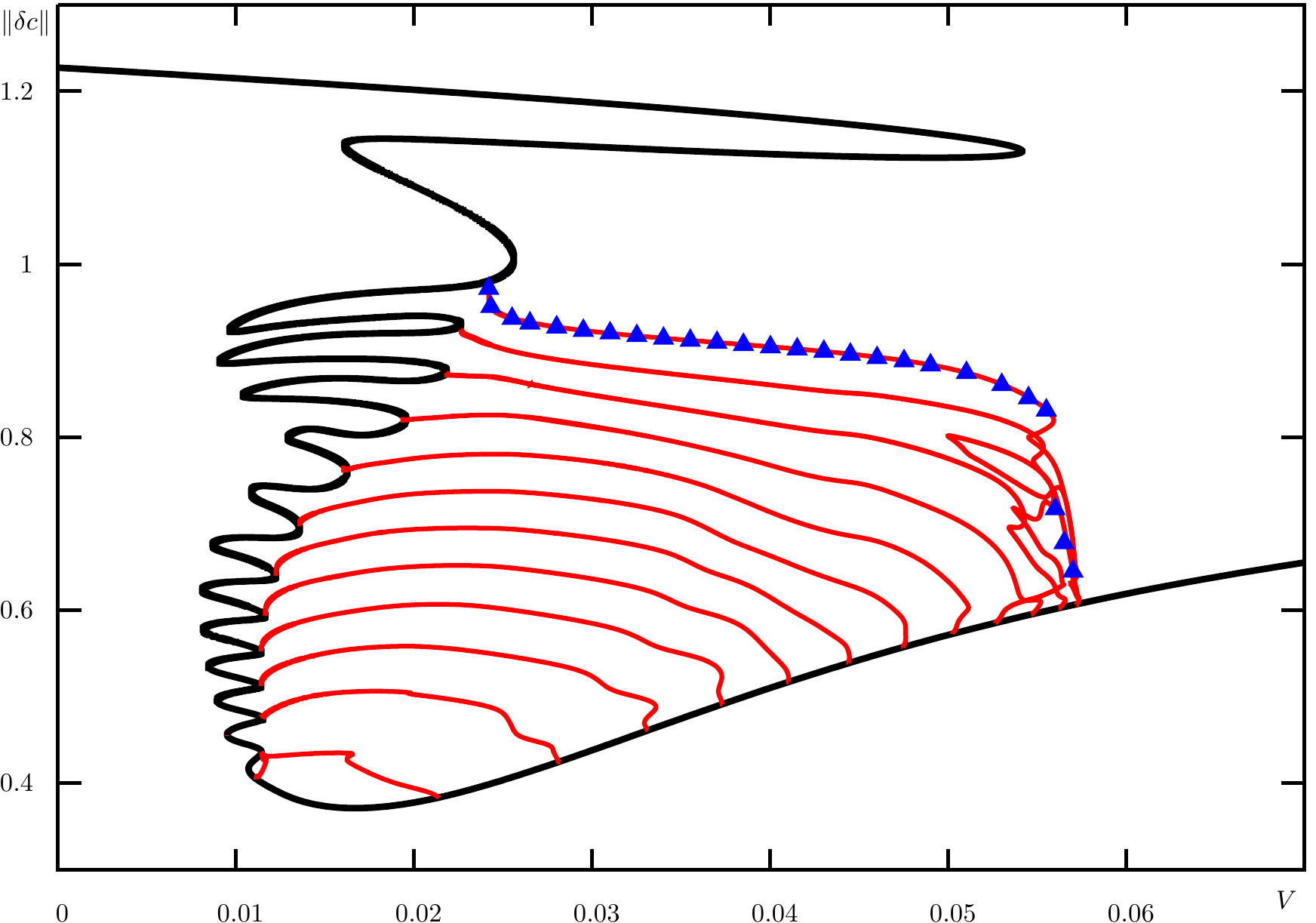}
\caption{Shown is the (time-averaged) $\lvert \delta c \rVert$ norm of steady and
  time-periodic solutions of Eq.~(\ref{eq:reducedexpl}) in dependence
  of the dimensionless plate velocity $V$ for $L=100$ and all the
  remaining parameters as in Fig.~\ref{fig:numeration}. The thick
  black solid and thin red solid lines represent steady states
  (homogeneous deposition) and time-periodic solutions (line
  deposition), respectively, as obtained by numerical path
  continuation. The blue triangles correspond to time-periodic
  solutions obtained by direct numerical simulation.  Selected zooms
  showing important details are given in Figs.~\ref{fig:ends}
  and~\ref{fig:PD}.}
  \mylab{fig:allperiodic}
\end{figure}
\begin{figure}
\centering
\includegraphics[width=0.45\hsize]{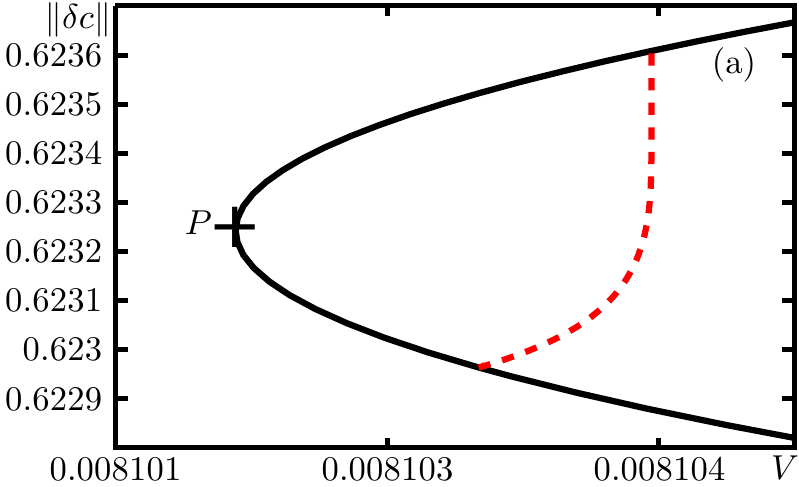}\hspace{0.05\hsize}
\includegraphics[width=0.45\hsize]{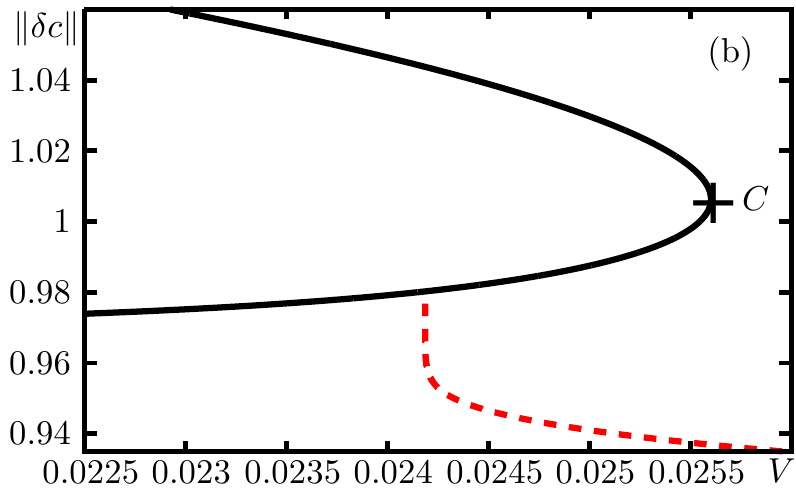}\\
\includegraphics[width=0.45\hsize]{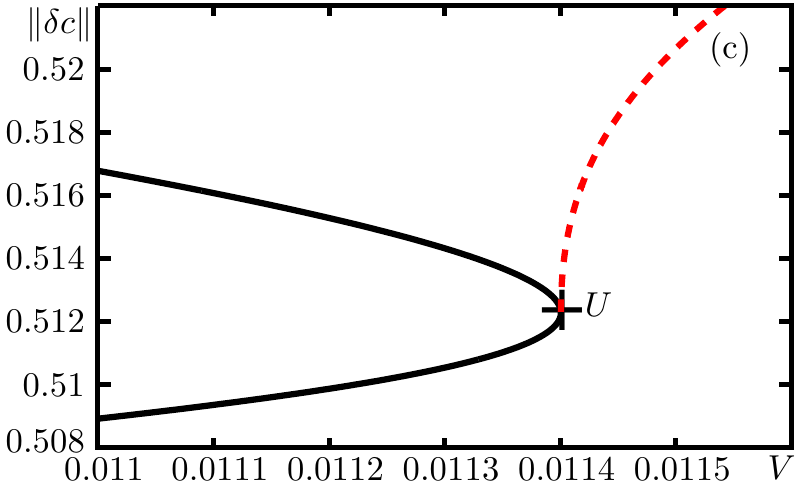}\hspace{0.05\hsize}
\includegraphics[width=0.45\hsize]{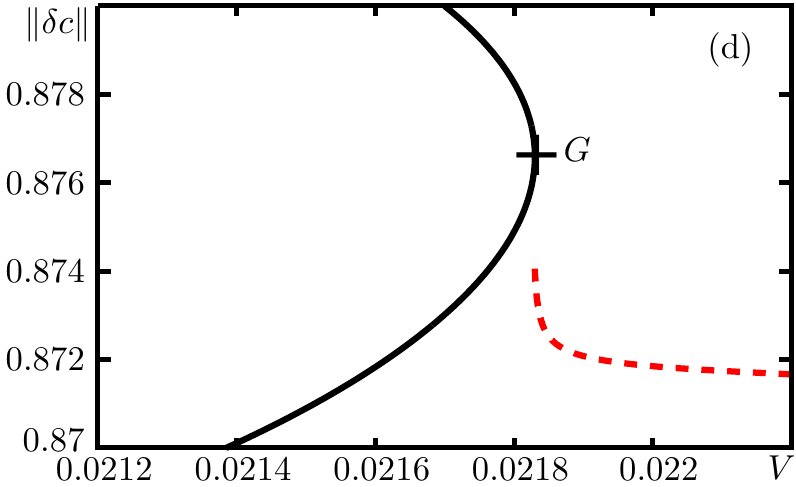}\\
\includegraphics[width=0.45\hsize]{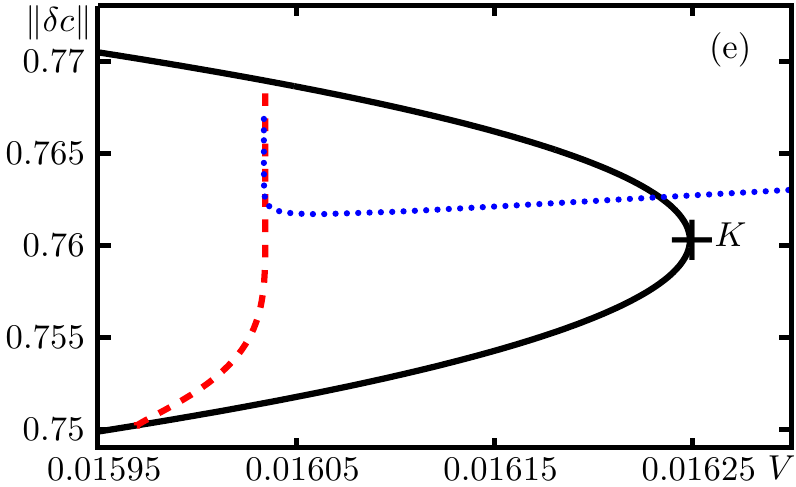}\hspace{0.05\hsize}
\includegraphics[width=0.45\hsize]{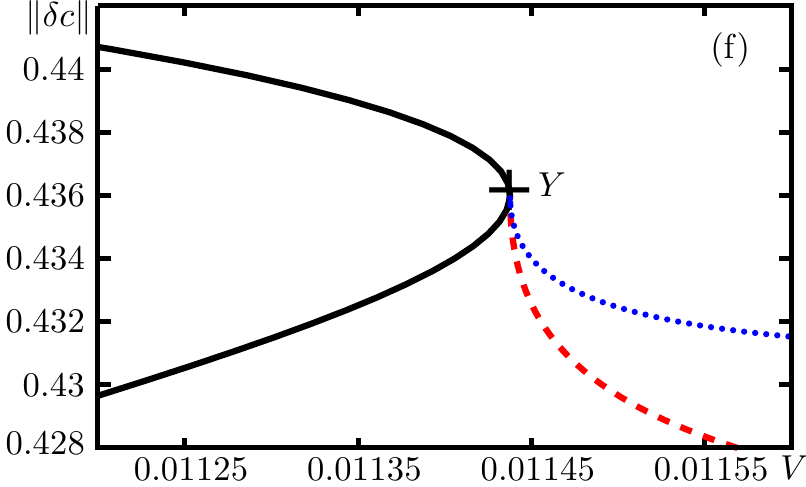}
\caption{\label{fig:ends}Close-ups of the bifurcation diagram
  Fig.~\ref{fig:allperiodic} zooming in on qualitatively different
  endpoints of branches of time-periodic solutions that
 involve global bifurcations. Panels (a) to (f) correspond to cases (i1), (i2),
 (ii1), (ii2), (iii) and (iv) discussed in the main text.}
\end{figure}
\begin{figure}
\centering
\includegraphics[width=.495\textwidth]{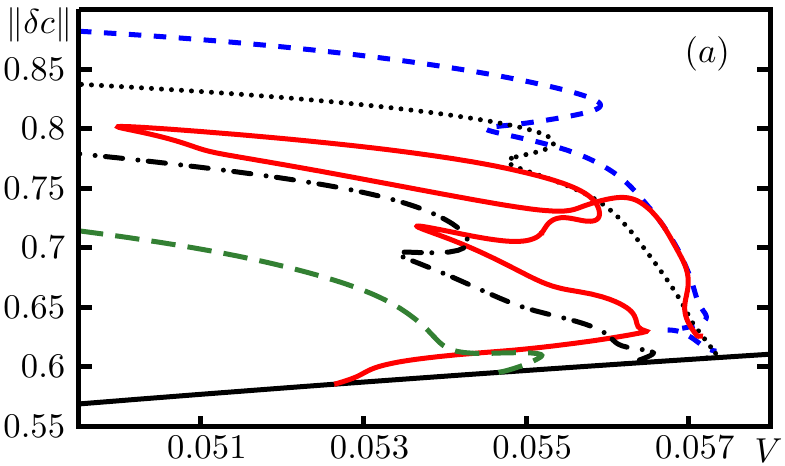}
\includegraphics[width=.495\textwidth]{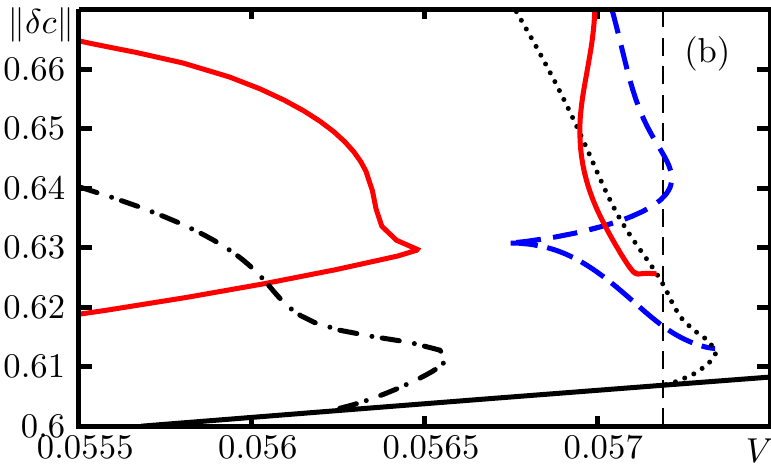}
\caption{\label{fig:PD} Close-ups of the bifurcation diagram
  Fig.~\ref{fig:allperiodic} zooming in on period doubling
  bifurcations and complex reconnections in the region
  $0.55<V<0.58$. Shown are all branches of periodic solutions that
  emerge from Hopf bifurcation (HB) 26 (dotted), HB~25 (dot-dashed), HB~24 (green long dashed),
  HB~23 (red solid line), as well as a secondary branch (blue short dashed)
  emerging from the first period doubling bifurcation of the HB~26
  branch. The stable time-periodic solutions observed in the direct
  time simulations lie on this secondary branch.  Note that all the
  Hopf bifurcations at high velocities are subcritical. The vertical
  dashed line in panel (b) marks the velocity $V=0.5719$ that is used for the time
  simulation shown in Fig.~\ref{fig:simtraject} and indicates the
  coexisting solutions at this velocity.}
\end{figure}

To complete the bifurcation diagram Fig.~\ref{fig:numeration}, i.e.,
the case of $L=100$ that is also investigated in
Ref.~\cite{KGFT2012njp}, we continue all the branches of time-periodic
solutions emerging at the Hopf bifurcations that we call 'primary Hopf
branches', as well as selected branches of time-periodic solutions
emerging from the primary Hopf branches in period-doubling
bifurcations. The latter we call 'secondary Hopf branches'. As a
result we find the harp-like structure presented in Fig.~\ref{fig:allperiodic}.
Each primary Hopf branch that starts
either sub- or supercritically at HB~15 to HB~26, i.e., right of the
point where the norm is minimal, eventually turns towards smaller $V$,
continues to lower $V$ ('spanning the harp') and finally ends in a
global bifurcation on the snaking part of the branch of steady
states. The global bifurcation is either a saddle-node infinite period
(sniper) bifurcation or a homoclinic bifurcation
\cite{Stro94,Kuzn10}. There are several secondary Hopf branches that
connect a period doubling bifurcation of a primary Hopf branch (e.g.,
solid red and dashed blue branches in Fig.~\ref{fig:PD}) either to a
global bifurcation on the snake (e.g., branch that connects the first
PD of the HB~26 branch to a homoclinic bifurcation near fold $C$), or
to a primary HB (e.g., branch that connects the second PD of HB~26 to
HB~24, see Fig.~\ref{fig:PD}~(a)) or to another period doubling
bifurcation of a primary Hopf branch (see Fig.~\ref{fig:LVHomSni}~(b)
below).
Beside these extended branches, there are (very) short branches that
emerge at HB~1 to HB~13 close to folds~$D, F, I-N, P, R, T, V, X$ and
$Z$ and connect to a global bifurcation nearby (beyond the next
  fold of the snaking structure).  The following list summarises the
  features of the five different scenarios in which global
  bifurcations are observed at low $V$ for branches that emerge from
  Hopf bifurcations or period-doubling bifurcations at higher $V$.
\begin{itemize}
\item[(i)] The branch of time-periodic solutions monotonically
  approaches a vertical asymptote at some critical $V$ and ends in a
  homoclinic bifurcation at an unstable steady state on the snake.
  One may distinguish sub-cases (i1) and (i2) with the branch of
  time-periodic states approaching the homoclinic bifurcation from the
  left and from the right, respectively. The case (i1) are the short
  branches emerging from HB~1, 2, 4, 6 and 8-13 that lie completely
  within one bend of the snake. For an example see
  Fig.~\ref{fig:ends}~(a) (note the tiny $V$ range). The case (i2) are
  the short branches emerging from HB~3, 5~and~7 and the long branch
  that emerges from a period doubling bifurcation of the HB~1 branch
  (see Fig.~\ref{fig:ends}~(b)). The latter branch consists mainly of
  the stable time-periodic solutions observed by direct time
  simulations \cite{KGFT2012njp}.
\item[(ii)] The branch of time-periodic solutions monotonically
  approaches a vertical asymptote at a critical $V$ that corresponds
  to a saddle-node bifurcation of the snake of steady states and ends
  in a sniper bifurcation.  One may distinguish sub-cases (ii1) and
  (ii2) with the branch of time-periodic states approaching the
  saddle-node bifurcation from above in the $V-||\delta c||$ plane and
  from below, respectively. The case (ii1) are the branches emerging
  from HB~16-21. For an example see Fig.~\ref{fig:ends}~(c). The case
  (ii2) are the branches coming from HB 15, 25, and 26 (see
  Fig.~\ref{fig:ends} (d)).
\item[(iii)] Two branches of time-periodic solutions monotonically
  approach seemingly the same vertical asymptote at a critical $V$
  where they end in two homoclinic bifurcations. This occurs for the
  branches that emerge from HB~3 and~24 and end at the branch between
  folds $H$ and $I$ as well as for the branches that emerge from HB~5
  and~22 and end at the branch between folds $J$ and $K$ (see
  Fig.~\ref{fig:ends} (e). For our calculations the respective two
  homoclinic bifurcations occur at values of $V$ that are identical up
  to 4 significant digits.
\item[(iv)] As (iii), but the two branches end in a sniper bifurcation
  at a fold (or in a sniper bifurcation and a homoclinic one very
  close nearby. This occurs for the branches that emerge from HB~14
  and~15 (see Fig.~\ref{fig:ends} (f)). For our calculations the two
  bifurcations occur at values of $V$ that are identical up to 5
  significant digits. 
  As we will explain later, this scenario is related to a
  codimension 2 bifurcation that creates the cases (i) and (ii).
\end{itemize}

The information regarding the branches of time-periodic solutions
contained in Figs.~\ref{fig:allperiodic} to~\ref{fig:PD} goes well
beyond Ref.~\cite{KGFT2012njp} and sheds a new light on results
obtained there. In particular, a weakly nonlinear analysis of the
branch emerging at the Hopf bifurcation at the highest velocity
(HB~26) led to the conclusion, that this HB is subcritical. However,
in Ref.~\cite{KGFT2012njp} it was also observed through time
simulations, that the phase space trajectories of the system close to
HB~26, where only one pair of complex conjugate modes becomes
unstable, are very complicated: First, they exhibit a long phase of
slow exponential growth (outwards spiral on a plane in phase
space). After this initial phase, deviations from this planar dynamics
are observed in the form of large excursions from the spiral which are
then caught by a higher dimensional structure on which the trajectory
eventually returns to the plane and begins to follow the spiral,
again. This dynamics is not simply periodic and might be chaotic.  An
attempt to understand this behaviour as an example of a scenario
described by Ovsyannikov and Shilnikov \cite{OS_MathUSSR_92}, which
involves a stable and an unstable pairs of complex conjugate
eigenvalues failed because it was found that there is a strong interaction
of the unstable mode and more than three of the stable modes
\cite{KGFT2012njp} . 

\begin{figure}
\centering
\includegraphics[width=.8\textwidth]{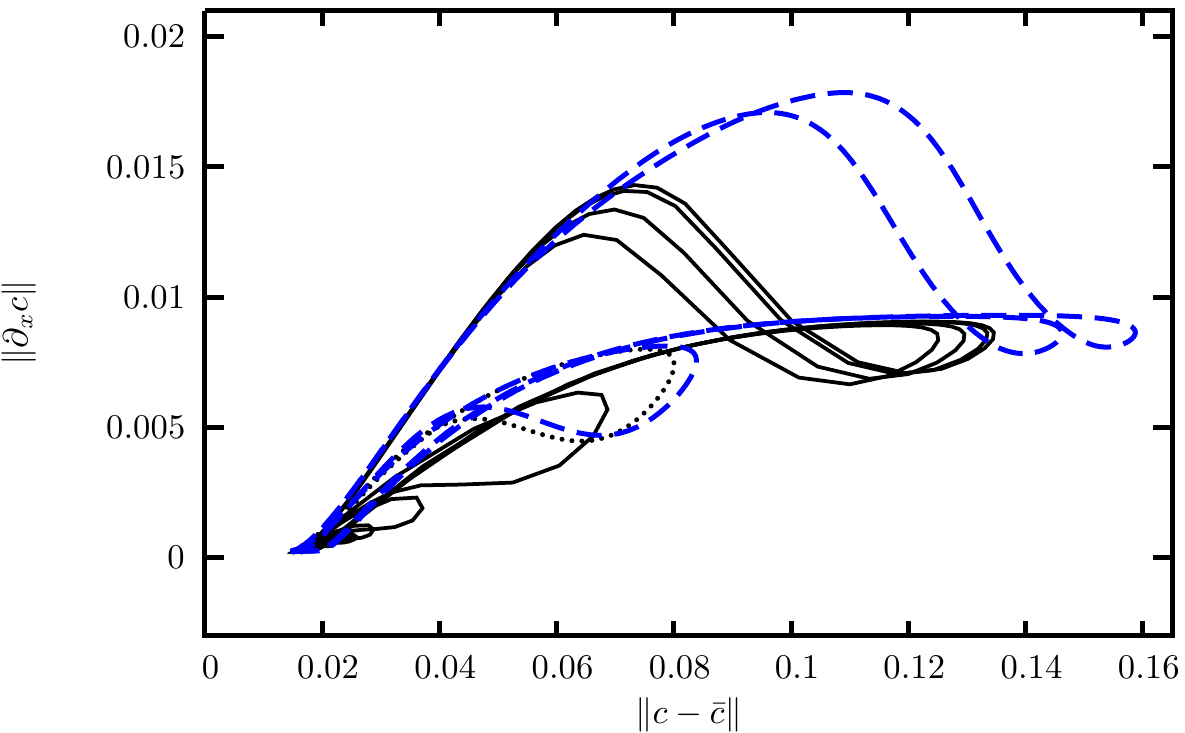}
\caption{\label{fig:simtraject} Shown are representations of various
  time-dependent solutions on the plane spanned by the norms of
  $c(x,t)-\bar{c}$ and of $\partial_x c(x,t)$, where $\bar{c}$ is the
  mean value of $c$. A trajectory obtained by direct numerical
  simulation is given as a thin solid line while the time-periodic
  branches obtained by numerical continuation are shown in the same
  linestyles used in Fig.~\ref{fig:PD}. The value of the plate
  velocity is $V=0.05719$.}
\end{figure}

In view of the close-ups of Fig.~\ref{fig:allperiodic} in
Fig.~\ref{fig:PD}, the complicated trajectories and the higher
dimensional structure which attracts the dynamics can be understood as
a consequence of the existence of several (up to 12) stable and
unstable time-periodic solutions at the same velocity
value. The majority of these solutions is unstable.  As the
system spirals away from the unstable fixed point (steady state
solution) it feels the attracting and repelling presence of the
various coexisting time-periodic solutions nearby. This can actually
be seen in the results of direct numerical simulations performed at a
value of $V$ close to HB~26: Once the initial amplification has become
large enough, the system behaviour shows clear similarities to the
time-periodic branches. This is indicated in Fig.~\ref{fig:simtraject} that
shows phase plane representations of a simulated trajectory and of the various
coexisting time-periodic branches at $V=0.05719$. One notes that
although the trace of the simulated time evolution is similar to the
time-periodic solutions, it does not approach any of them closely, indicating
that they act as repellers.  Even in very long simulations, the trajectory does
not settle onto any of the stable time-periodic solutions, but eventually
returns to a phase space region close to the unstable fixed point where the
exponential amplification sets in again. Note that the role of a large number
of repellers in the transition to chaos has been analysed, e.g., in
Ref.~\cite{EcMe1999pre} for channel flow.

In the present section we have analysed the bifurcation structure for fixed
$L=100$ employing $V$ as main control parameter. To further understand how the
observed harp-like structure emerges we next investigate how the bifurcation
structure changes with our second control parameter $L$.
%
\subsection{Simplification of bifurcation structure with decreasing domain size $L$}\mylab{sec:res-small}
%
\begin{figure}
\centering
\includegraphics[width=0.32\textwidth]{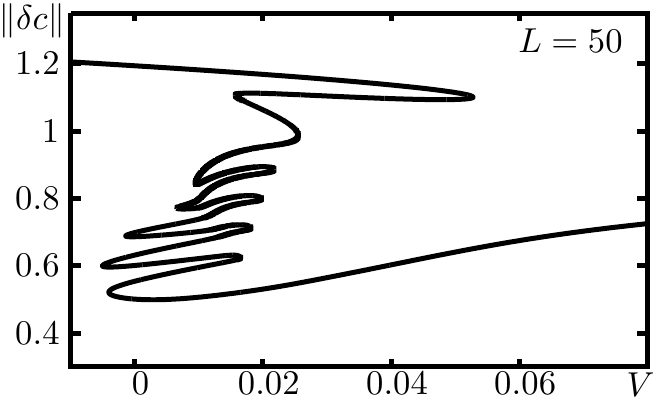}
\includegraphics[width=0.32\textwidth]{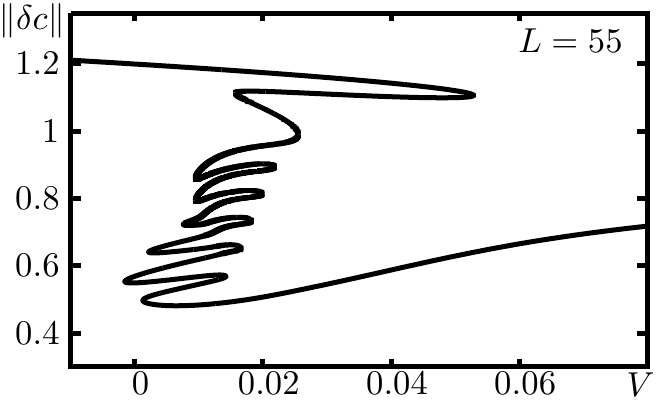}
\includegraphics[width=0.32\textwidth]{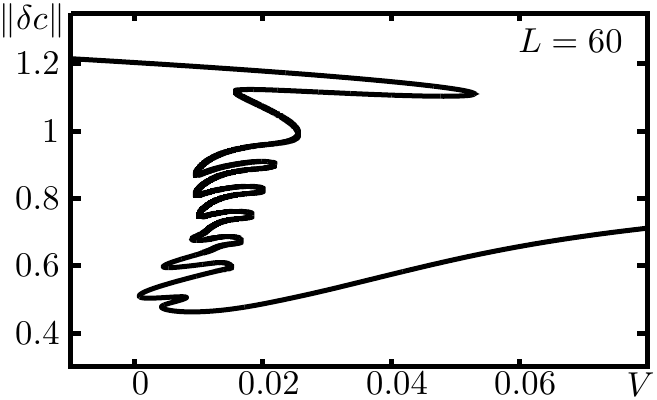}
\includegraphics[width=0.32\textwidth]{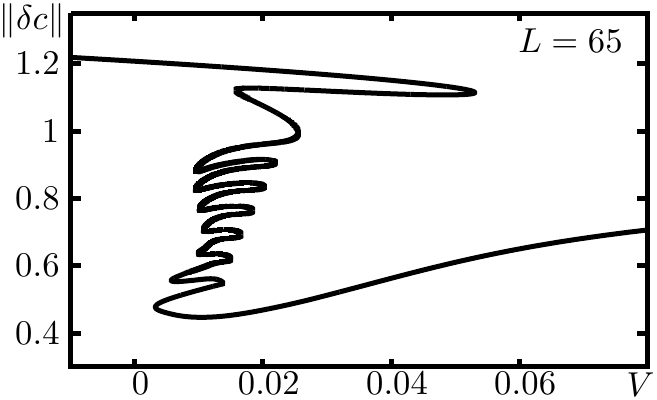}
\includegraphics[width=0.32\textwidth]{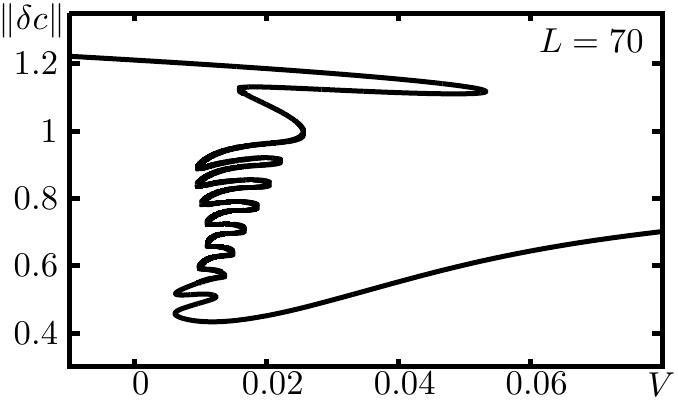}
\includegraphics[width=0.32\textwidth]{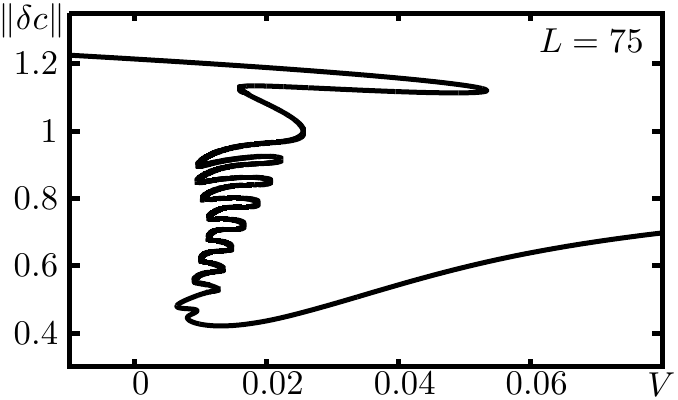}
\includegraphics[width=0.32\textwidth]{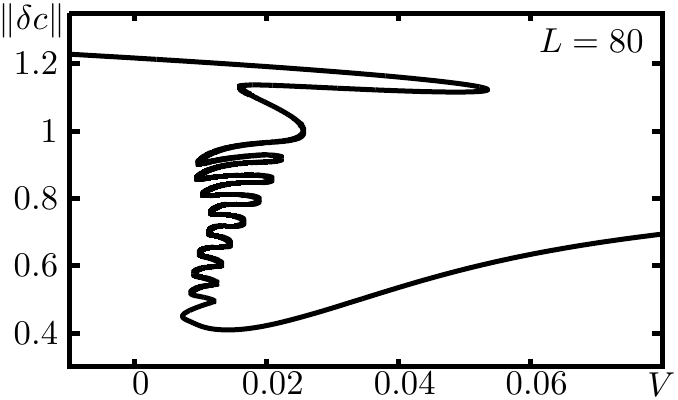}
\includegraphics[width=0.32\textwidth]{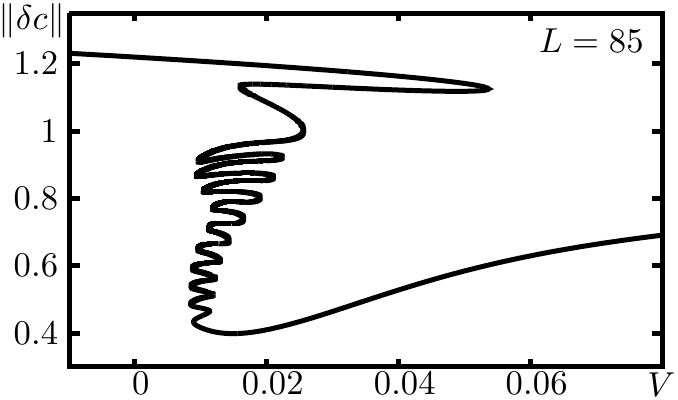}
\includegraphics[width=0.32\textwidth]{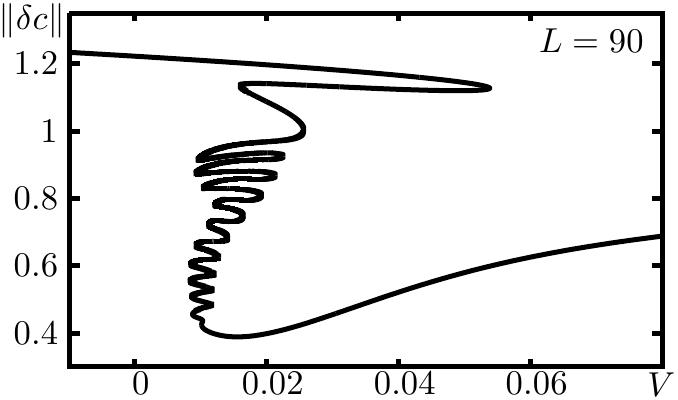}
\caption{\mylab{fig:smallerLs} Shown is the branch of steady solutions
  for equally spaced values of the domain size from $L=50$ to
    $L=90$ as indicated in the individual panels (a) to (i). All the
  remaining parameters are as in Fig.~\ref{fig:numeration}. The
  occurring simplification with decreasing $L$ is discussed in the
  main text. Fig.~\ref{fig:LVHomSni} show close-ups of the final bends
  of the snaking structure including the time-periodic branches for
  $L=90$ to $L=100$.
}
\end{figure}

Up to here we have completed the bifurcation diagram
Fig.~\ref{fig:numeration}, i.e., in the case of $L=100$ partially
given in Ref.~\cite{KGFT2012njp}, by a systematic continuation of the
branches of time-periodic solutions emerging at all of the 26 HB.
This has allowed for a classification of possible endings of branches
of time-periodic solutions corresponding to the behaviour at the onset
of line deposition. Overall, the analysis indicates that our
particular system has a complicated bifurcation structure that
involves saddle-node and Hopf bifurcations of steady states, period
doubling bifurcations of time-periodic solutions and various global
bifurcations, namely sniper and homoclinic bifurcations.

These observations are in line with observations in the literature for
a wider class of related systems that show deposition or depinning
phenomena. For other line deposition phenomena, similar bifurcations
are hypothesised based on time simulations in
Refs.~\cite{FrAT2011prl,FrAT2012sm}. In particular,
Ref.~\cite{FrAT2012sm} discusses in which way such line deposition
patterns may be seen as resulting from depinning transitions, thereby
relating the phenomena to the depinning of driven droplets on
heterogeneous substrates
\cite{ThKn2006njp,BKHT2011pre,HTHB2012el,VFFP2013prl}, rotating
cylinders \cite{Thie2011jfm}, or to the depinning of interface undulations from
the advancing tip of an air finger in a liquid-filled channel \cite{PHGJ2012pf,Thompson14}.
In these related systems depinning can also occur through Hopf, sniper or
homoclinic bifurcations. This will be further discussed in the conclusion.

The main physical ingredients of our particular system are a first
order phase transition and a lateral driving (towards the right hand
side, see Fig.~\ref{fig:sketch}) that keeps the system permanently out
of equilibrium. Close to the left boundary (the meniscus side), the
underlying energy is a symmetric double well. As the concentration on
the boundary is fixed to a value close to one of the minima of the
energy, the corresponding phase is preferred.  This changes to the
right of the critical position $x_\mathrm{s}$ where the double well
potential is tilted such that the other phase is preferred.

Considering our main influences, we already know that in the case
without driving $V=0$ one obtains a standard system that evolves
towards equilibrium, i.e., in the right hand part it approaches the
high-concentration phase promoted by the tilted double well
potential. Out of equilibrium one finds at low [large] $V$ a
homogeneous deposition of the preferred phase [the boundary phase]. At
intermediate $V$ the interplay of driving and phase dynamics results
in stripe deposition.  

One can also expect that no structure formation will occur if the
intrinsic length scale of the patterning $\ell$ is larger than
the considered domain size $\tilde{L}$. This implies that the rich
structure of the bifurcation diagram Fig.~\ref{fig:allperiodic} should
vanish step by step if $L=\tilde{L}/\ell$ is decreased.  In the
following, we focus on this stepwise simplification and, in order to
observe it, we compute bifurcation diagrams for decreasing
dimensionless domain size $L$ that represents the ratio of dimensional
domain size and intrinsic patterning length scale. In the alternative
scaling discussed in note \cite{note_scaling}, one would keep $L$
fixed and reduce the quench depth $a$.

We focus on the range $50 \le L \le 100$ as for smaller $L$ the distance $x_s$
between the boundary and the position where the tilt of the potential
part of the free energy changes becomes important and renders the
emerging picture non-generic. 
Figure \ref{fig:smallerLs} shows the branch of steady solutions for various $L$
in the considered range. One notices that
with increasing $L$, the snaking structure becomes larger, i.e., more and more
pairs of saddle-node bifurcations emerge in subsequent hysteresis bifurcations,
adding wiggles to the branch that then widen with increasing $L$. In this way,
one passes from 12 folds at $L=50$ to 26~folds at $L=100$. 
This is of course expected, since with each turn of the snake, a bump
is added to the solution profile until the part of the snake with the
lowest norm corresponds to a state in which the complete domain to the
right of the substrate-mediated condensation threshold is filled with bumps (see
Fig.~\ref{fig:profiles}~(e)). With increasing $L$, more bumps fit into
the domain, thus prolonging the snake.

\begin{figure}
\centering
\includegraphics[width=0.7\textwidth]{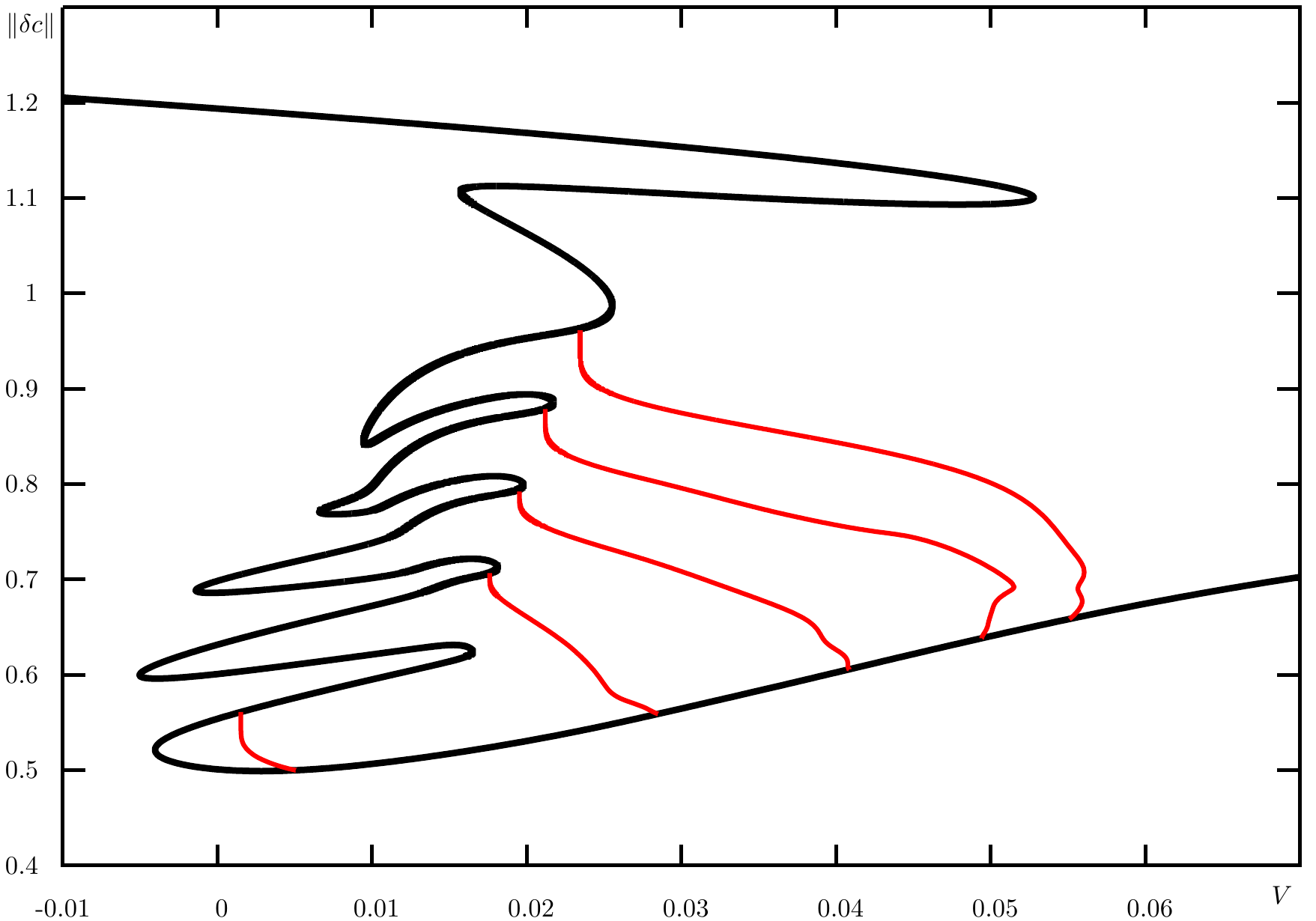}
\caption{\mylab{fig:allperiodicL50} Shown is the (time-averaged) $\lVert \delta c \rVert$ norm of steady and
  time-periodic solutions of Eq.~(\ref{eq:reducedexpl}) in dependence
  of the dimensionless plate velocity $V$ for $L=50$ and all the
  remaining parameters as in Fig.~\ref{fig:numeration}. The thick
  black solid and thin red solid lines represent steady states
  (homogeneous deposition) and time-periodic solutions (line
  deposition), respectively, as obtained by numerical path
  continuation. Note, that the reconnection of the rightmost branches that is
  observed in the case $L=100$ (see Fig.~\ref{fig:allperiodic}) is not observed
  at $L=50$.} \end{figure}

However, from the continuation results shown in
Fig.~\ref{fig:allperiodic} one knows that the snaking structure is
closely linked to the branches of time-periodic solutions. This raises
the question what happens to the branches observed at $L=100$ when the
snaking structure shrinks with decreasing domain size $L$. Figure
\ref{fig:allperiodicL50} shows the bifurcation diagram with steady and
time-peridic solution for $L=50$. One can see a smaller version of the
harp-like structure that we observed for $L=100$ in Fig.~\ref{fig:allperiodic}.
With decreasing $L$, as more and more wiggles disappear from the branch of
steady solutions, more and more branches of time periodic solutions vanish as well
and the structure shown in Fig.~\ref{fig:allperiodic} evolves to the
structure shown in Fig.~\ref{fig:allperiodicL50}.

We observe a generic sequence of transitions that eliminate branches of
time-periodic solutions and the related global and local bifurcations
in parallel to the elimination of folds of the snake when $L$ is
decreased. We explain the mechanism using the example of the
time-periodic branches of lowest norm in Fig.~\ref{fig:allperiodic},
in particular the ones emerging from HB~13-16. All the involved
solution branches are shown for eight values of $L$ between 92 and 100
in Fig.~\ref{fig:LVHomSni} while Fig.~\ref{fig:peroverVhopfzoom} shows
the loci of the involved Hopf, homoclinic, sniper and saddle-node
bifurcations in the parameter plane spanned by $V$ and $L$.

\begin{figure}
\centering
\includegraphics[width=0.45\textwidth]{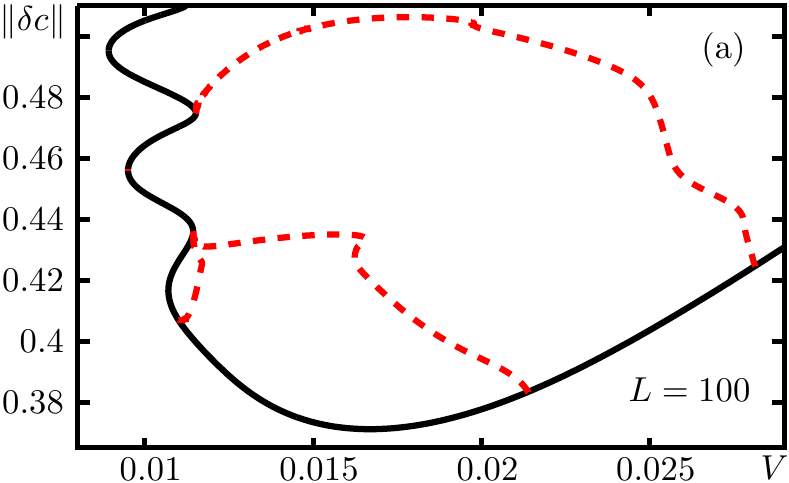}\hspace{5mm}
\includegraphics[width=0.45\textwidth]{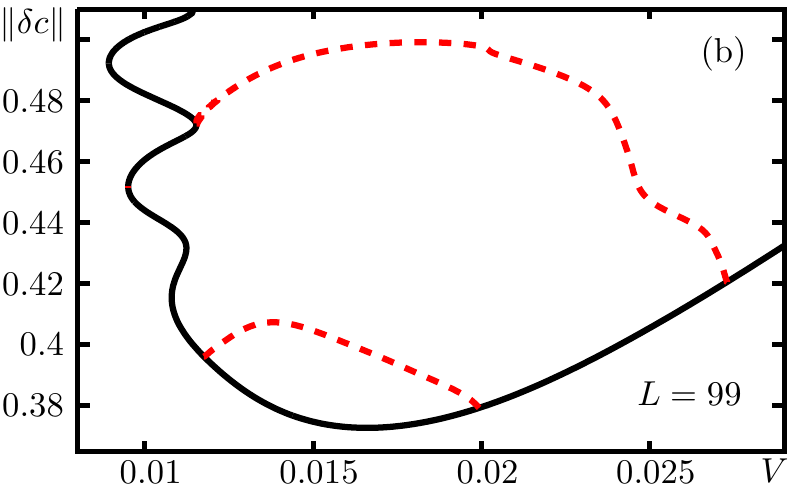}
\includegraphics[width=0.45\textwidth]{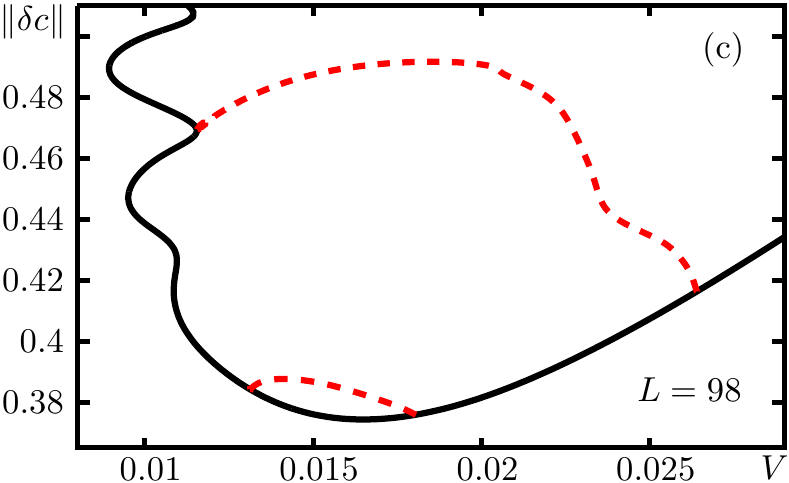}\hspace{5mm}
\includegraphics[width=0.45\textwidth]{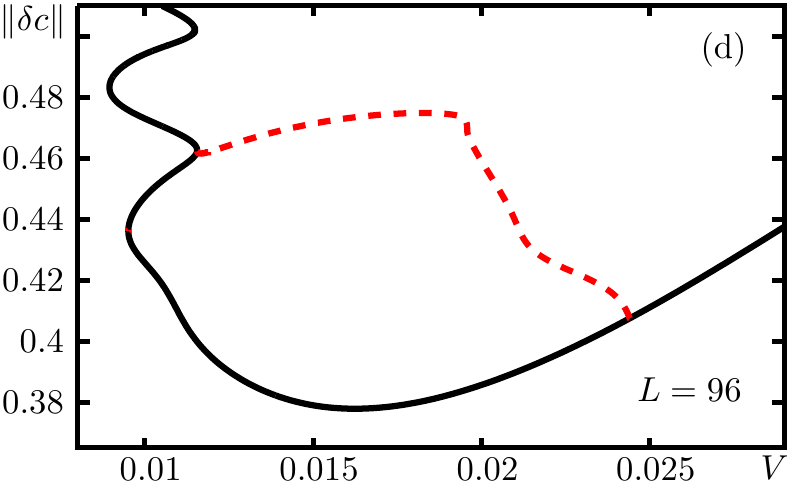}
\includegraphics[width=0.45\textwidth]{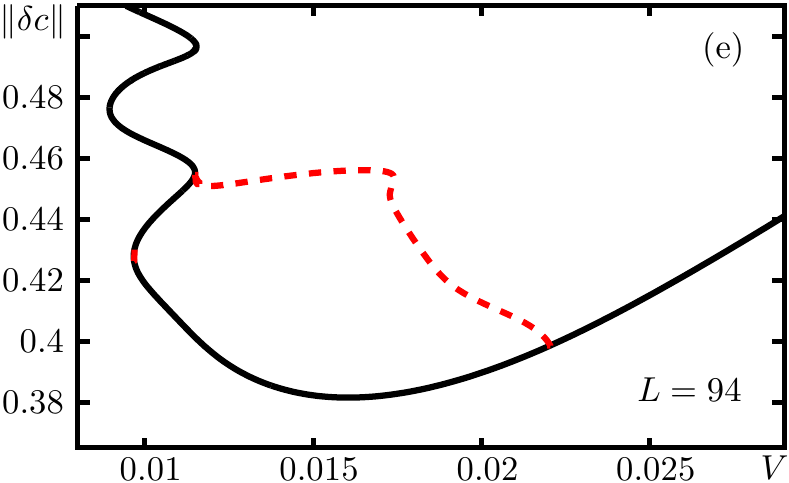}\hspace{5mm}
\includegraphics[width=0.45\textwidth]{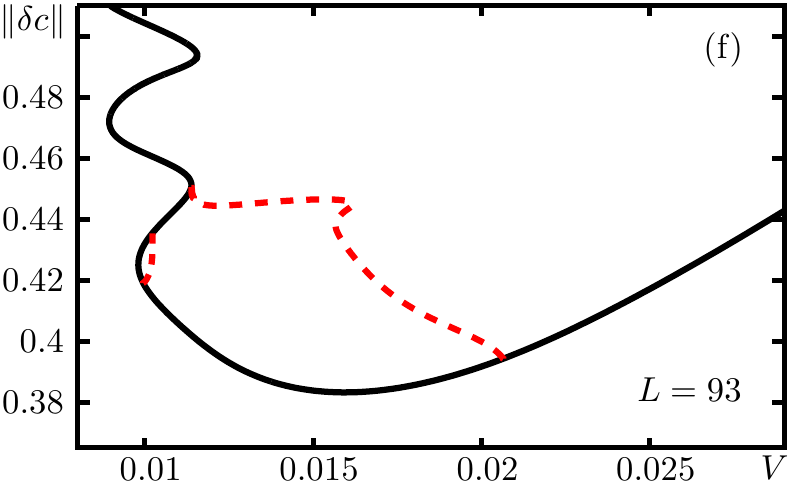}
\includegraphics[width=0.45\textwidth]{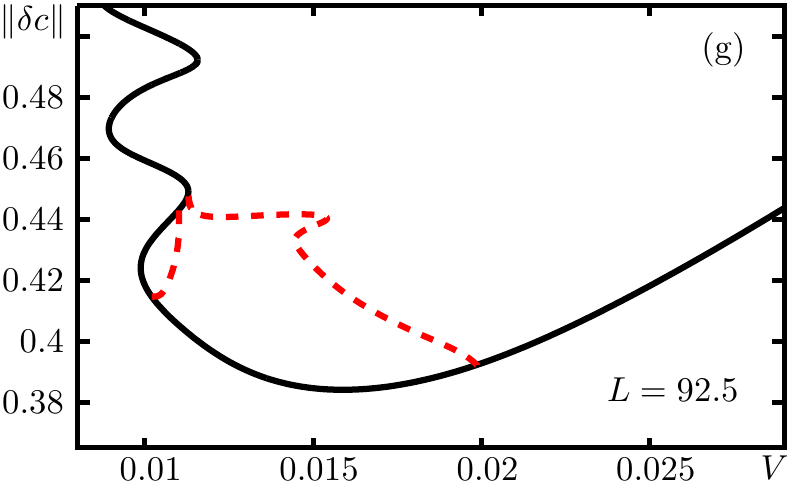}\hspace{5mm}
\includegraphics[width=0.45\textwidth]{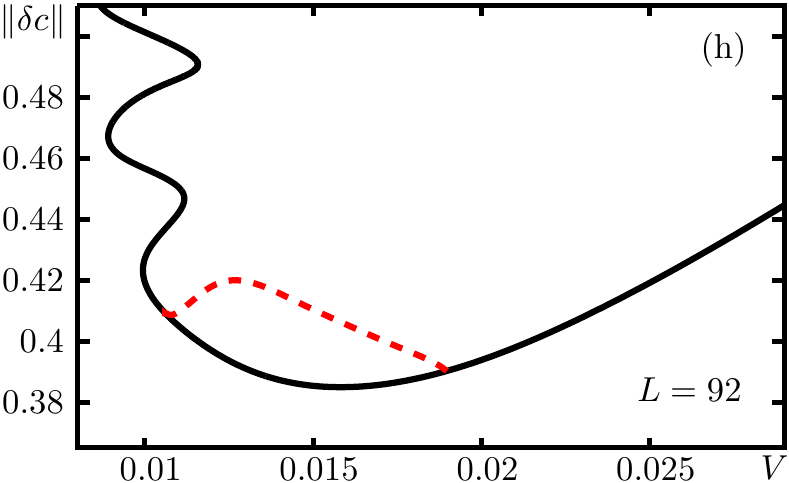}
\caption{\mylab{fig:LVHomSni} Shown are close-ups of the lower part of
  the snaking structure for
  various values of domain size $L$ from 100 to 90 as indicated in the
  individual panels (a) to (h).  The thick black solid and red dashed
  lines represent steady states (homogeneous deposition) and
  time-periodic solutions (line deposition), respectively, as obtained
  by numerical path continuation.}
\end{figure}

\begin{figure}
\centering
\includegraphics[width=0.8\textwidth]{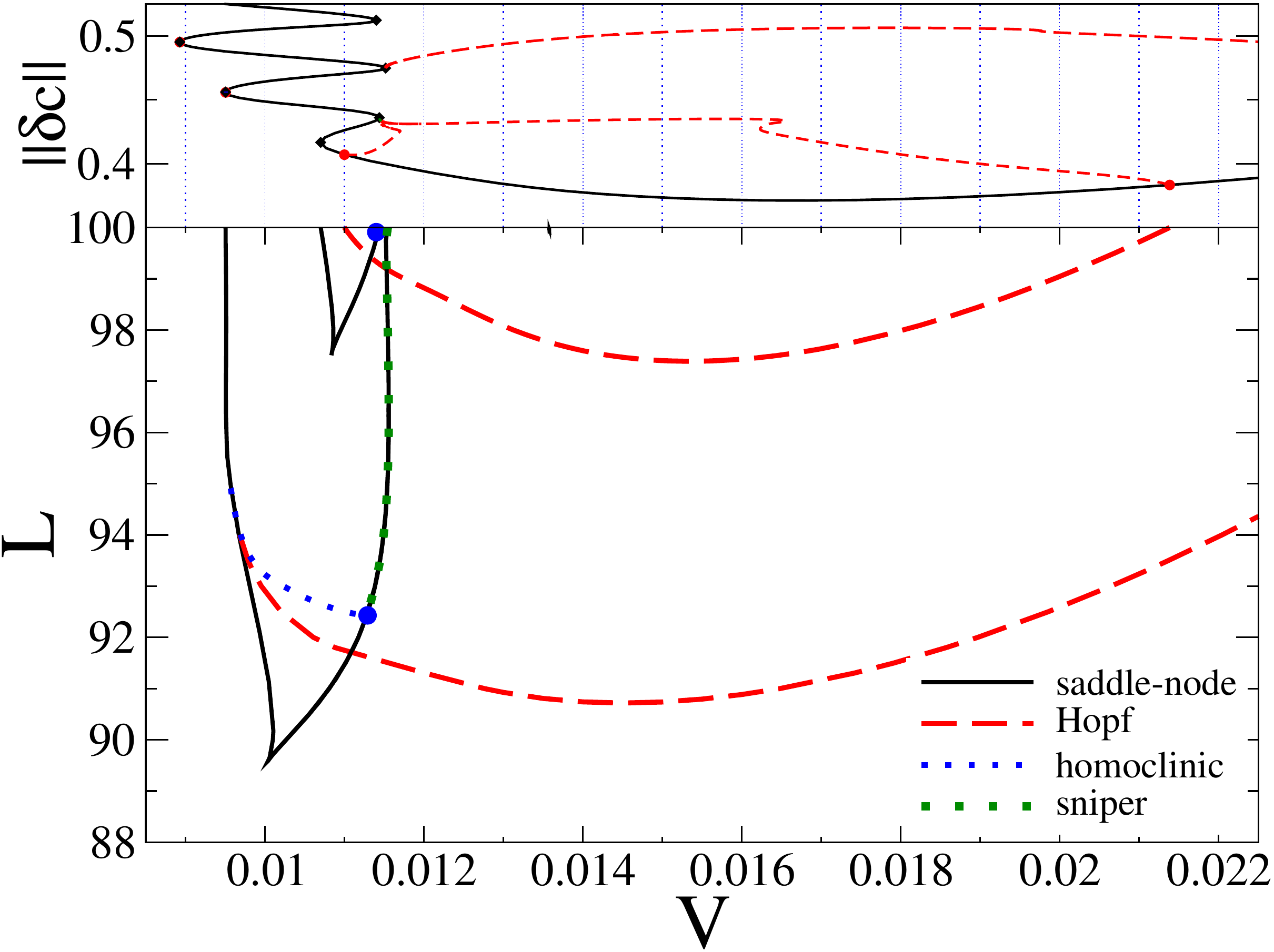}
\caption{\mylab{fig:peroverVhopfzoom}Loci of the folds $V$ to $Z$
  (black solid lines), the Hopf bifurcations 13 to 16 (red dashed
  lines), and of the homoclinic bifurcation where the primary branch
  ends that emerges from HB~14 (dotted line).  Blue filled circles
  indicates where a branch of time-periodic solutions collides with
  the branch of steady states creating a homoclinic and a sniper
  bifurcation. Above the filled circle a sniper bifurcation coincides
  with the black solid line of the respective fold.  }
\end{figure}

We describe the transitions with increasing $L$, i.e., the various
bifurcations and branches are \textit{created} subsequently. At $L=90$
none of the Hopf bifurcations~13-16 and none of the related
time-periodic branches exist, the lowest pair of saddle-node bifurcations of
steady states has just been created in a hysteresis bifurcation at $L\approx
89.5$ (lower black solid line in Fig.~\ref{fig:peroverVhopfzoom}). Slightly
below $L=91$ one observes a codimension 2 bifurcation that occurs close to
the minimum of the curve of steady states. As a result a pair of Hopf bifurcations (HB~13
and~16) appears together with the branch of time-periodic solutions that
connects them. The creation of two HB corresponds to the minimum of the lower
red dashed curve in Fig.~\ref{fig:peroverVhopfzoom} that indicates the loci of
the two HB. With increasing $L$, the two HB move apart and the left (lower $V$)
HB moves towards the lowest fold of the snake; see the bifurcation diagram at
$L=92$ in Fig.~\ref{fig:LVHomSni}~(h).
At about $L=92.5$, the branch of time-periodic solutions that connects
HB~13 and~16 breaks up into two disconnected branches. This happens
when the branch of time-periodic solutions develops a protrusion that
eventually collides with the unstable steady state slightly beside or
directly at the lowest right fold ($W$). This codimension 2
bifurcation is marked by the lower blue filled circle in
Fig.~\ref{fig:peroverVhopfzoom}. The bifurcation diagram in
Fig.~\ref{fig:LVHomSni}~(g) is at a slightly larger $L$, the two
global bifurcations are already separated. The new left branch of
time-periodic solutions coming from HB~13 at lower $V$ now ends in a
homoclinic bifurcation that moves towards smaller $V$ (dotted line in
Fig.~\ref{fig:peroverVhopfzoom}) while the right branch of
time-periodic solutions, coming from HB~16 at larger $V$, terminates
in a sniper bifurcation at fold $W$. Further
increasing $L$, the left branch becomes
shorter and shorter, and practically disappears into fold $X$, while
the right branch becomes longer as HB~16 moves continuously towards
larger velocities and the other end remains at fold $W$ (see Figs.~\ref{fig:LVHomSni} (a)-(f) and
Fig.~\ref{fig:peroverVhopfzoom}). In the course of this process, there
are also two folds created on the branch of time-periodic solutions
that are again annihilated at about $L=95$.
Note that the HB and the homoclinic bifurcation that remain very close
to the left fold (on opposite sites) could, in principle annihilate (and be
recreated) in Bogdanov-Takens bifurcations \cite[see section~8.4. of]{Kuzn10},
but whether this actually happens, e.g., between $L\approx93$ and $L\approx97$
we are not able to decide on the basis of the present numerical accuracy.

Then the entire process is repeated: At about $L=97.5$, a pair of
saddle-node bifurcations of steady states is created (upper black
solid line in Fig.~\ref{fig:peroverVhopfzoom}) and a pair of Hopf
bifurcations (HB~14 and~15) appears together with the branch of
time-periodic solutions that connects them (minimum of the upper red
dashed line in Fig.~\ref{fig:peroverVhopfzoom}). The two HB move apart
and the left one approaches fold $Z$, and two global bifurcations are
created when the branch of time-periodic solutions connects to the
steady state branch close to fold $Y$ (see the bifurcation diagrams in
Fig.~\ref{fig:LVHomSni}~(a)-(c). By this mechanisms, new turns of the
snake and new branches of time-periodic solutions can be generated for
arbitrary system sizes.
%
\section{Discussion and conclusion}
\mylab{sec:conc}
%
We have explored the bifurcation structure of a modified Cahn-Hilliard equation
that describes a system that may undergo a first order phase transition and is
kept permanently out of equilibrium by a lateral driving. This form of the
equation was introduced in Ref.~\cite{KGFT2012njp} as a model for the
deposition of stripe patterns of surfactant molecules through Langmuir-Blodgett
transfer, i.e., the transfer of a surfactant monolayer from the free
surface of a water bath onto a solid substrate that is drawn out of
the bath at a controlled velocity \cite{RS_ThinSolidFilms_92}. The
pattern formation results from a first order structural liquid-liquid
phase transition that is triggered by the intermolecular interactions
with the solid plate that become effective when the monolayer gets
close to the substrate, i.e., in the contact line region. The effect
is called substrate-mediated condensation
\cite{note_SMC,RS_ThinSolidFilms_92}. The modified Cahn-Hilliard
equation represents a reduced model that captures the main features of
the patterning as observed before with a hydrodynamic long-wave
description in terms of coupled evolution equations for the film
height profile and the surfactant concentration
\cite{KGFC_Langmuir_10} that themselves may be seen as a special case of a
gradient dynamics description for the coupled evolution of the density
of a layer of an insoluble surfactant and of the film thickness
profile of the underlying film of liquid \cite{ThAP2012pf}.

The modified Cahn-Hilliard model captures the main physical ingredients that
are the first order phase transition and the lateral driving that 
together with the pinning boundary condition on the meniscus side keeps the system
permanently out of equilibrium. The system is biased towards one of the two
phases only beyond a threshold position that represents the position of the
contact line region in the physical system. The state in the boundary region on
the mensicus side is determined by the boundary condition. In the literature
other forms of such Cahn-Hilliard-type models are studied
\cite{Kr_PRE_09,FW_PRE_09,FW_PRE_12}. In particular, in these works a front
that switches the system between a stable one-phase and an unstable two-phase
region of the phase diagram is moved through a fixed domain of given mean
concentration. However, the onset of pattern deposition is not studied in a way
that could be directly compared to our results. In particular, no bifurcations
between steady states and time-periodic states (in the frame moving with the
imposed front speed) are discussed.

Our analysis has shown that the snaking of steady states (observed in
Ref.~\cite{KGFT2012njp}) that correspond to various fronts between a
homogeneous low concentration region close to the left domain boundary
and a homogeneous state far away from the bath is intertwined with a
large number of branches of time-periodic solutions that emerge from
Hopf bifurcations or period doubling bifurcations and end mainly in
global bifurcations (sniper and homoclinic) but sometimes also in
other period doubling or Hopf bifurcations. Overall the various
branches form a harp-like structure (see Fig.~\ref{fig:allperiodic}).

Beside the particular study of the bifurcation diagram for a fixed
value of the domain size we have analysed how the bifurcation structure
simplifies if one decreases the dimensionless domain size, i.e.,
increases the typical intrinsic length scale of pattern formation
related to the first order transition. We have shown that for decreasing
$L$ the rich bifurcation structure vanishes step by step through a
number of codimension 2 bifurcations, namely, annihilation of pairs of
homoclinic and sniper bifurcations, annihilations of pairs of period
doubling bifurcations, and annihilations of pairs of Hopf
bifurcations.  Beside these transitions involving branches of
time-periodic solutions, the front snaking structure of steady
solutions reduces by annihilations of pairs of saddle-node
bifurcations.

The intriguing question of how the complicated bifurcation scenario at
large domain size emerges is not only relevant for the system of
Langmuir-Blodgett transfer but for a wider class of systems showing
deposition and depinning phenomena. It has been argued in
Ref.~\cite{FrAT2012sm} that the onset of line deposition, i.e., the
emergence of branches of time-periodic solutions may be seen as
resulting from a depinning transition: When a homogeneous layer is
deposited on the moving plate the concentration profile is steady in
the laboratory frame and one may say that the concentration profile is
pinned. However, when lines are deposited, in the laboratory frame
they move away from the bath, i.e., the lines are depinned from the
meniscus. This implies that one should not only expect to find
similarities between the present study and other studies of the
bifurcation structure behind deposition patterns as, e.g.,
Refs.~\cite{FrAT2011prl,FrAT2012sm,DoGu2013el} (see also review
\cite{Thie2014acis}), but see such similarities as well in studies of
driven droplets depinning from substrate heterogeneities
\cite{ThKn2006njp,BKHT2011pre,HTHB2012el,VFFP2013prl} and rotating
cylinders \cite{Thie2011jfm}; studies of driven clusters of
interacting colloidal particles depinning in heterogeneous pores
\cite{PAST2011pre}; or studies of bubbles advancing in a channel with
a central geometrical constraint, where the bubbles develop
undulations that are steady in the laboratory frame \cite{PHGJ2012pf,Thompson14}.
In the frame moving with the uniformly advancing tip of the bubble the
undulations depin from the tip similar to the depinning of a
deposition pattern from the contact line region. A related
  effect is also observed in abrasive wawterjet cutting, a technique of cutting
  material with a focused beam of abrasive particles accelerated by water that is moved across the material
  \cite{Hashish_ExpMech_1988,Folkes_JMatProcTech_2009}. There is a model based on a Kuramoto-Sivashinsky-type
  equation that shows a transition from smooth cuts (analogeous to
  our homogeneous deposition) to ripples and striations patterns (analogeous to
  our line deposition) \cite{FRDH_PRL_00,Radons2005}. In this system the undulations represent
  structures that depin from the moving cutting jet.

In all these systems the transition from steady (pinned) states to
time-periodic (depinned) states is also found to occur through sub- or
supercritical Hopf, sniper or homoclinic bifurcations. Although
transitions between different scenarios where observed, in particular,
in Refs.~\cite{ThKn2006njp,BKHT2011pre,FrAT2012sm,HTHB2012el}, the
occurring codimension 2 bifurcations involving time-periodic states 
have not yet been identified.

For each of the mentioned systems one can find a physically meaningful
control parameter whose variation brings the system into a part of the
parameter space where no large amplitude structures exist, i.e., the
system becomes either steady at all values of the imposed driving
(here the plate velocity $V$) or only allows for small amplitude 'surface waves' (as in
the Kuramoto-Sivashinsky limit of the convective Cahn-Hilliard or thin
film equation). However, although depinning and deposition patterns
are widely occurring, to our knowledge no systematic study exists for any of
the mentioned systems that investigates the emergence/vanishing of the
rich bifurcation structure related to depinning or pattern deposition
that should characterize these system. Therefore, the
results obtained here will be of some relevance to other similar
systems as they provide a sound basis for comparison.

The present study is limited in two important ways.  On the one hand,
it is a study of solutions to a partial differential equation 
with one spatial dimension only, i.e., the described involved
bifurcation structure only captures solutions that are invariant
w.r.t.\ translations along the contact line and is, e.g., not able to
capture the transition between the deposition of stripes parallel and
perpendicular to the contact line observed in
Refs.~\cite{KGFC_Langmuir_10,KGF_PRE_11,KGFT2012njp} for
Langmuir-Blodgett transfer and in the Cahn-Hilliard-type models where
quench fronts are moved through the system \cite{Kr_PRE_09,FW_PRE_12}.
Further large-scale numerical bifurcation studies will be needed to
systematically explore the full solution space for out-of-equilibrium
amended Cahn-Hilliard-type models allowing for a breaking of the
transversal geometry in analogy to such studies for depinning droplets
\cite{BeHT2009el,BKHT2011pre}. However, also the latter need to be
extended to understand how the entire bifurcation structure vanishes
if one goes from partially wetting liquids to completely wetting
liquids.

We remark that a future study should also further investigate the
influence of the boundary conditions, in particular, on the meniscus
side. The need of boundary conditions that allow for a controlled
through-flow of material as in the present system is a common
complication in dragged plate problems. In the case of simple liquids,
recently a central manifold reduction was employed to rigorously
derive an asymptotic series for the meniscus side that can then be
used to construct boundary conditions that do not introduce an
additional domain-size dependence \cite{TsGT2014epje}. Such an
approach has still to be developed for complex fluids.

On the other hand, the present study is entirely based on numerical
path continuation while the analytical results obtained in
Ref.~\cite{KGFT2012njp} are entirely based on linear and weakly
nonlinear analyses. As it is not to be expected that the uncovered
complex bifurcation scenario of the full PDE can be treated
analytically, there is a strong need for further simplified models
based on ordinary differential equations (ODE) that are able to
reproduce the main features described here. We expect that they need
to be at least four-dimensional. Note, that the authors of
Ref.~\cite{BKHT2011pre} were able to use selected results of the ODE
studies in Refs.~\cite{Sigg2003jfm,KrOl2004n} to explain certain
aspects of the depinning behaviour of droplets, however, also there
the 'mapping' remains rather incomplete. We hope that our numerical
studies of the bifurcation structure of deposition and depinning
processes can motivate the development of generic ODE models for this
class of processes.

\acknowledgments

The authors are grateful to the Newton Institute in Cambridge, UK, for
its hospitality during their stay at the programme ``Mathematical
Modelling and Analysis of Complex Fluids and Active Media in Evolving
Domains`` where part of this work was done.
MHK acknowledges the support by the Human Frontier Science Program (Grant
RGP0052/2009-C). This publication is based in part on work supported by Award
No.~KUK-C1-013-04, made by the King Abdullah University of Science and Technology
(KAUST), and LabEX ENS-ICFP: ANR-10-LABX-0010/ANR-10-IDEX-0001-02 PSL*.


%
\bibliography{lbtransfer} 

\begin{thebibliography}{10}
\bibitem{note_SMC}
The first order structural transition is between the liquid-expanded phase and
  the liquid-condensed phase of the surfactant layer. One speaks of a
  ''substrate-mediated condensation'' if for a thick sub-layer of water the
  surfactant layer remains in the liquid-expanded one phase region while on a
  very thin water layer the presence of the solid substrate underneath lowers
  the free-energy of the condensed state and thus effectively facilitates
  condensation \cite{RS_ThinSolidFilms_92}. The effect is well studied for the
  phospholipid DPPC (dipalmitoylphosphatidylcholine), but should in principle
  occur for any surfactant that exhibits such a first order transition.
\bibitem{note_scaling}
For instance, chosing the scales $\Gamma=(\tilde\mu/\kappa)^{1/3}$,
  $\tau=\sigma/M\kappa^{2/3}\tilde\mu^{4/3}$, and
  $\ell=\sigma^{1/2}/\kappa^{1/6}\tilde\mu^{1/3}$ one obtains
  Eq.~(\ref{eq:scales}) with $M$, $\sigma$, $\tilde\mu$ and $\kappa$ set to one
  and without the tildes. The non-dimensional plate velocity is still $V=\tilde
  V \tau/\ell$ and the four remaining nondimensional parameters are $V$, $a$,
  $L$ and $c_0$.
\end{thebibliography}
%
%
\end{document}